\documentclass[aps,pra,notitlepage,nofootinbib]{revtex4-1}
\usepackage[T1]{fontenc}
\usepackage[latin9]{inputenc}
\usepackage{amscd}
\usepackage{amsmath}
\usepackage{esint}
\usepackage[unicode=true,
 bookmarks=false,
 breaklinks=false,pdfborder={0 0 1},backref=section,colorlinks=false]
 {hyperref}
\usepackage{breakurl}

\makeatletter

\usepackage{amscd}
\usepackage{breakurl}

\usepackage{amscd}
\usepackage{breakurl}

\usepackage{bbm}

\usepackage{amsfonts}
\usepackage{times}
\usepackage{graphicx}
\usepackage{braket}
\usepackage{amscd}

\newcommand{\g}{\mathfrak{g}}
\renewcommand{\U}{\mathfrak{U}}
\renewcommand{\sl}{\mathfrak{sl}}
\newcommand{\su}{\mathfrak{su}}
\newcommand{\so}{\mathfrak{so}}
\renewcommand{\C}{\mathbb{C}}
\newcommand{\R}{\mathbb{R}}
\newcommand{\id}{\mathbbm{1}}
\newcommand{\N}{\mathbb{N}}
\newcommand{\ad}{\text{ad}}
\newcommand{\Tr}{\text{Tr}}

\newcommand{\z}{z}
\renewcommand{\L}{L}

\newcommand{\D}{{\cal D}}
\renewcommand{\l}{\ell}
\newcommand{\kk}{{\cal K}}
\renewcommand{\r}{r}

\makeatother

\begin{document}

\title{Non-commutative Fourier transform for the Lorentz group via the Duflo
map}

\author{Daniele Oriti}

\affiliation{Max Planck Institute for Gravitational Physics (Albert Einstein Institute),
~~~\\
 Am Mühlenberg 1, 14476 Potsdam-Golm, Germany ~~~\\
 {} ~~~\\
 Arnold-Sommerfeld-Center for Theoretical Physics, Ludwig-Maximilians-Universität,
~~~\\
 Theresienstrasse 37, D-80333 München, Germany, EU}

\author{Giacomo Rosati}

\affiliation{INFN, Sezione di Cagliari, Cittadella Universitaria, 09042 Monserrato,
Italy }
\begin{abstract}
We defined a non-commutative algebra representation for quantum systems
whose phase space is the cotangent bundle of the Lorentz group, and
the non-commutative Fourier transform ensuring the unitary equivalence
with the standard group representation. Our construction is from first
principles in the sense that all structures are derived from the choice
of quantization map for the classical system, the Duflo quantization
map. 
\end{abstract}
\maketitle

\section{Introduction}

Physical systems whose configuration or phase space is endowed with
a curved geometry include for example point particles on curved spacetimes
and rotor models in condensed matter theory, and they present specific
mathematical challenges in addition to their physical interest. In
particular, many important results have been obtained in the case
in which their domain spaces can be identified with (non-abelian)
group manifolds or their associated homogeneous spaces.

Focusing on the case in which the non-trivial geometry is that of
configuration space, identified with a Lie group, this is then reflected
in the non-commutativity of the conjugate momentum space, whose basic
variables correspond to the Lie derivatives acting on the configuration
manifold, and that can be thus identified with the Lie algebra of
the same Lie group. At the quantum level, this non-commutativity enters
heavily in the treatment of the system. For instance, it prevents
a standard $L^{2}$ representation of the Hilbert state space of the
system in momentum picture, and may lead to question whether a representation
in terms of the (non-commutative) momentum variables exists at all.
Further, if such a representation could be defined, one would also
need a generalised notion of (non-commutative) Fourier transform to
establish the (unitary) equivalence between this new representation
and the one based on $L^{2}$ functions on the group configuration
space.

The standard way of dealing with this issue is indeed to renounce
to have a representation that makes direct use of the non-commutative
Lie algebra variables, and to use instead a representation in terms
of irreducible representations of the Lie group, i.e. to resort to
spectral decomposition as the proper analogue of the traditional Fourier
transform in flat space. In other words, instead of using some necessarily
generalised eigenbasis of the non-commutative momentum (Lie algebra)
variables, one uses a basis of eigenstates of a maximal set of commuting
operators that are functions of the same. This strategy has of course
many advantages, but it prevents the more geometrically transparent
picture, more closely connected to the underlying classical system,
that dealing directly with the non-commuting Lie algebra variables
would provide.

\
 In recent times, most work on these issues has been motivated by
quantum gravity, where they turned out to play an important role.
This happened in two domains. The first is effective (field theory)
models of quantum gravity based (or inspired) by non-commutative geometry
\cite{NC-effective}. Such models have attracted a considerable interest
because they offer a framework for much quantum gravity phenomenology
\cite{QGphen}. From the physical point of view, these models are
based on the hypothesis that some form of spacetime non-commutativity
\cite{NC-effective} constitutes a key remnant, at a macroscopic,
continuum level, of more fundamental quantum gravity structures, and
that one key way in which this manifests itself is by relaxing \cite{NC-locality}
(or making relative \cite{RelLocPRL,relative-locality}) the usual
notion of locality on which effective field theory (and most spacetime
physics) is based. At the mathematical level, spacetime non-commutativity
in these models is the conjugate manifestation of curvature in momentum
space (and thus in phase space) \cite{JurekFre-kfield1e2}. In most
of them, moreover, this curved momentum space is modeled by a non-abelian
Lie group (the phase space is therefore its cotangent bundle) or homogeneous
spaces constructed from it, hence the relevance of the non-commutative
Lie algebra representation and of the non-commutative Fourier transform
for Lie groups (see for example \cite{MajidOeckl-kFuorier,LukKosMas-kfield,GACMajid-kMink}).
The second domain is that of non-perturbative quantum gravity proper,
more specifically the broad area comprising group field theory \cite{GFT},
spin foam models \cite{SF} and loop quantum gravity \cite{LQG}.
Here again the cotangent bundle of a Lie group plays a prominent role:
it is the classical phase space of the variables associated to the
(edges of the) lattices, which replace continuum spacetime manifolds
and represent the basic interaction processes of the fundamental degrees
of freedom of all these formalisms. The interpretation of these algebraic
variables have, in turn, a discrete geometric interpretation, as discrete
gravity connection (the group variables) and conjugate discrete metric
variables (the Lie algebra elements). Once again, the Lie algebra
representation and the non-commutative Fourier transform allow a representation
of the formalisms in terms of more geometrically transparent data
\cite{NC-GFT,NC-LQG}; this, in turn, facilitates model building as
well as the extraction of effective macroscopic physics.

\
 In this article we are not going to deal with these physical aspects,
although we will discuss them briefly. We focus instead on the formal
aspects of the non-commutative Fourier transform. Among the many issues
motivated by its intrinsic mathematical interest, it is important
to clarify its general properties for arbitrary Lie groups and its
explicit construction in the challenging case of non-compact ones.
The mathematical foundations of the Lie algebra representation and
of the non-commutative Fourier transform have been explored in very
general terms, for example, in ~\cite{GuOrRaNCFT}. A key result
of that work has been to show how both the star-product and the non-commutative
plane waves that characterize fully such algebra representation and
its equivalence with the other representations of the underlying quantum
system, are uniquely determined by the initial choice of quantization
map that defines the same quantum system starting from its classical
phase space and algebra of observables. Moreover, this has been shown
for arbitrary semi-simple locally compact Lie groups. Thus, while
the explicit constructions reported in the same paper only dealt with
the compact group $SU(2)$, the same procedure can be straightforwardly
extended to the non-compact case. This is what we do in this paper,
focusing on the Lorentz group $SL(2,\mathbb{C})$ because of its general
physical interest and because it plays in particular a key role in
quantum gravity, in the context of spin foam models, group field theories
as well as effective non-commutative spacetime models. The quantization
map we use, which is the key technical input of the whole construction,
is the Duflo map \cite{duflo}. This has already found some applications
in quantum gravity \cite{AlekseevDuflo,FreMajDuflo,SahlThiemDuflo,SahlThiemDufloPRL,NouiPerezPranzDuflo,VitaleDuflo},
and it had been used in ~\cite{GuOrRaNCFT} for the compact $SU(2)$
case. For our present purposes, the mathematical features of this
quantization map that makes it a preferred choice are its generality
(it also can be applied to any semi-simple Lie algebra) and its characterizing
property of mapping the subalgebra of invariant functions (under the
group action) on the classical phase space faithfully to a subalgebra
of invariant operators in the quantum theory. This is mathematically
desirable, but also physically crucial, in our opinion.

\
 The plan of the paper is as follows. In the next section, we summarize
the results of ~\cite{GuOrRaNCFT}, for general Lie groups, which
form the basis for our subsequent analysis and construction for the
Lorentz group. In section ~\ref{sec:dufloLorentz-algebra} we use
the Duflo map to construct the Lie algebra representation of a quantum
system whose classical phase space is the cotangent bundle of the
Lorentz group, and then, in section ~\ref{sec:NCfourier}, using
the same quantization map, we obtain the non-commutative Fourier transform
that ensure its equivalence with the more traditional group representation,
constructing explicitly the corresponding non-commutative plane waves.
These are the key new results of this paper. We also offer a quick
comparison with related results in the literature, concerning the
definition and application of non-commutative methods for similar
systems, both at the mathematical level and in quantum gravity contexts.
Moreover, we show how our results generalise to the case of homogeneous
spaces of the Lorentz group. Finally, in the last section, we show
the application of our tools to the computation of the propagator
of a single non-relativistic point particle on the hyperboloid, as
an example.

\section{Background: Lie algebra representation and non-commutative Fourier
transform\label{sec:definitions}}

The mathematical foundations of the non-commutative Fourier transform
have been clarified in~\cite{GuOrRaNCFT}, and we base our present
work on that analysis. We summarize its results here.

\subsection{Classical phase space and its quantization}

We consider physical systems whose configuration space is a Lie group
$G$, and whose momentum space coincides with the dual $\g^{*}$ of
the Lie algebra $\g$ of $G$, giving the classical phase space as
the cotangent bundle $T^{*}G\cong G\times\g^{*}$. The classical algebra
of observables is thus the Poisson algebra ${\cal P}_{G}=\left(C^{\infty}\left(T^{*}G\right),\left\{ \cdot,\cdot\right\} ,\cdot\right)$,
such that 
\begin{equation}
\left\{ f,g\right\} \equiv\frac{\partial}{\partial X_{i}}{\cal L}_{i}g-{\cal L}_{i}f\frac{\partial g}{\partial X_{i}}+c_{ij}^{\ \ k}\frac{\partial f}{\partial X_{i}}\frac{\partial g}{\partial X_{j}}X_{k},
\end{equation}
for functions $f,g\in C^{\infty}\left(T^{*}G\right)$, Euclidean coordinates
$X_{i}$ on $\g^{*}\cong\mathbb{R}^{d}$ ($d:=\dim\left(G\right)$);
${\cal L}_{i}$ are Lie derivatives on $G$ (with respect an orthonormal
basis of right-invariant vector fields), and $c_{ij}^{\ \ k}$ are
the structure constants of the Lie algebra $\g$ $(\cong\g^{*})$
with $i,j,k=1,\dots,d$ (summation over repeated indexes is assumed).

The quantization of a maximal subalgebra ${\cal A}$ of ${\cal P}_{G}$
(for which the quantization is consistent) amounts to the definition
of an abstract operator $^{*}$-algebra $\U$, obtained from the classical
Poisson algebra. The quantization is defined through a quantization
map ${\cal Q}:\ {\cal A}\rightarrow\U$. The required properties of
the quantization map are as follows. First, ${\cal Q}\left(f\right)=:\hat{f}$
for all $f\in{\cal A}_{G}\subset C^{\infty}\left(G\right)$, ${\cal A}_{G}$
being the subalgebra of ${\cal A}\subset C^{\infty}\left(G\times\g^{*}\right)$
constant in the second argument, and ${\cal Q}\left(X_{i}\right)=:\hat{X}_{i}$,
satisfying 
\begin{equation}
\left[\hat{f},\hat{g}\right]=0,\quad\left[\hat{X}_{i},\hat{f}\right]=i\widehat{{\cal L}_{i}f}\in\U_{G},\quad\left[\hat{X}_{i},\hat{X}_{j}\right]=ic_{ij}^{\ \ k}\hat{X}_{k},\label{commOperators}
\end{equation}
for all $\hat{f},\hat{g}\in\U_{G}$ with $\U_{G}={\cal Q}\left({\cal A}_{G}\right)$.
Second, when restricted to functions ${\cal A}_{\g^{*}}\subset C^{\infty}\left(G\times\g^{*}\right)$
constant in the first factor, ${\cal Q}$ maps to a completion of
the universal enveloping algebra of $\g$, $\U_{\g^{*}}:={\cal Q}\left({\cal A}_{\g^{*}}\right)\cong\overline{U\left(\g\right)}$.
If we restrict to the space of polynomials in $\g^{*}$, then ${\cal Q}$
encodes the operator ambiguity coming from the non-commutativity of
the elements $\hat{X}_{i}$. The algebra $\U$ generated by $\hat{f}$
and $\hat{X}_{i}$ is then called the ``quantum (observable) algebra''
for $T^{*}G$.

$\U_{\g^{*}}$ is endowed with a natural Hopf algebra structure inherited
from the universal enveloping algebra $U$, with coproduct $\Delta_{\g^{*}}$,
counit $\epsilon_{\g^{*}}$ and antipode $S_{\g^{*}}$ 
\begin{equation}
\begin{gathered}\Delta_{\g^{*}}\left(\hat{X}_{i}\right)=\hat{X}_{i}\otimes\id+\id\otimes\hat{X}_{i},\\
\epsilon_{\g^{*}}\left(\hat{X}_{i}\right)=0,\qquad S_{\g^{*}}\left(\hat{X}_{i}\right)=-\hat{X}_{i}.
\end{gathered}
\end{equation}
An Hopf algebra structure in $\U_{G}$ can be obtained for exponential
Lie groups by considering canonical coordinates (of the first kind)
$k:\ G\rightarrow\g\cong\mathbb{R}^{d}$ defined through the logarithm
map, $k\left(g\right)=-i\ln\left(g\right)$, setting for the corresponding
operators $\epsilon_{G}\left(\hat{k}^{i}\right)=0$, $S_{G}\left(\hat{k}^{i}\right)=-\hat{k}^{i}$
and 
\begin{equation}
\Delta_{G}\left(\hat{k}^{i}\right)=\sum_{n=1}^{\infty}\sum_{\substack{k,l\in\N\\
k+l=n
}
}B_{p_{1}\cdots p_{k}q_{1}\cdots q_{l}}\hat{k}^{p_{1}}\cdots\hat{k}^{p_{k}}\otimes\hat{k}^{q_{1}}\cdots\hat{k}^{q_{k}},
\end{equation}
where $B_{p_{1}\cdots p_{k}q_{1}\cdots q_{l}}\in\R$ are the coefficient
of the Taylor expansion for the Baker-Campbell-Hausdorff formula $k^{i}\left(gh\right)={\cal B}\left(k\left(g\right),k\left(h\right)\right)^{i}$.
Defining the lift of $f\in{\cal A}_{G}$ onto the Lie algebra as $f_{k}\left(k\right)\equiv f\left(e^{ik}\right)$,
the coproduct can be extended to the corresponding $\hat{f}\in\U_{G}$
as $\Delta_{G}\left(\hat{f}\right)\equiv f_{k}\left(\Delta_{G}\left(\hat{k}^{i}\right)\right)$.
These definitions satisfy the requirements for the coproduct to be
induced by the group multiplication as $\Delta\left(f\right)\left(g,h\right)=f\left(gh\right)$,
as well as the group unit and inverse to induce counit and antipode,
$\epsilon\left(f\right)=f\left(e\right)$, $S\left(f\right)\left(g\right)=f\left(g^{-1}\right)$.

\subsection{The group and algebra representations}

In order to get a concrete realization of the abstract observable
algebra $\U$, one has to construct its representations, in which
the elements of the abstract algebra become operators acting on specific
Hilbert spaces.

For the system at hand, one can immediately define a ``group representation''
$\pi_{G}$ of $\U$ on the Hilbert space $L^{2}\left(G\right)$ of
square-integrable functions on $G$ with respect to the Haar measure
$dg$. This is done by diagonalizing all the operators $\hat{f}\in\U_{G}$
and setting the action of the $\hat{X}_{i}$ to be given by the Lie
derivatives, so that 
\begin{equation}
\begin{gathered}\left(\pi_{G}\left(\hat{f}\right)\psi\right)\left(g\right)\equiv f\left(g\right)\psi\left(g\right),\\
\left(\pi_{G}\left(\hat{X}_{i}\right)\psi\right)\left(g\right)\equiv i{\cal L}_{i}\psi\left(g\right),
\end{gathered}
\label{groupRep}
\end{equation}
for all $\psi\in C_{c}^{\infty}\left(G\right)$ where we restrict\footnote{\label{fn:compactness}This is necessary to ensure that $f\psi$ lies
in $C_{c}\left(G\right)$ for all $\psi\in C_{c}^{\infty}\left(G\right)$.
In general, since we are dealing with unbounded operators, for their
representation on a Hilbert space ${\cal H}$, their domain of definition
must be restricted to some dense subspaces of ${\cal H}$ such that
their images under the action of the operators are contained in ${\cal H}$.} the domain of $\pi_{G}$ to the sbspace of ${\cal A}_{G}$ of smooth
compactly supported functions $C_{c}\left(G\right)$ (dense in $L^{2}\left(G\right)$)
on $G$. The inner product for $\psi,\psi'\in L^{2}\left(G\right)$
is defined as usual as 
\begin{equation}
\left\langle \psi,\psi'\right\rangle _{G}\equiv\int_{G}dg\ \overline{\psi\left(g\right)}\psi'\left(g\right).\label{L2gNorm}
\end{equation}

That (\ref{groupRep}) is a representation of the quantum algebra
(\ref{commOperators}) is ensured by the fact that the Lie derivative
satisfies the Leibniz rule respect to the pointwise product of functions.
Such property can be expressed as a compatibility condition between
the coproduct $\Delta_{\g^{*}}$ of $\U_{\g^{*}}$ and the pointwise
product $m_{G}\ :\ f\otimes f'\rightarrow f\cdot f'$ for $f,f'\in C^{\infty}\left(G\right)$.
More in general if $\pi$ is a representation of $\U$ on a space
$\mathfrak{F}_{m}$ with multiplication $m\ :\ f\otimes f'\rightarrow f\cdot_{m}f'$,
the compatibility with the coproduct $\Delta$ can be expressed, for
an operator $\hat{T}$ in $\U$, by the commutative diagram 
\begin{equation}
\begin{CD}\mathfrak{F}_{m}\otimes\mathfrak{F}_{m}@>{m}>>\mathfrak{F}_{m}\\
@V{\pi\otimes\pi(\Delta(\hat{T}))}VV@VV{\pi(\hat{T})}V\\
\mathfrak{F}_{m}\otimes\mathfrak{F}_{m}@>{m}>>\mathfrak{F}_{m}
\end{CD}\label{productDiagram}
\end{equation}
expressing the identity $\pi(\hat{T})\circ m=m\circ\left(\pi\otimes\pi\right)(\Delta(\hat{T}))$.

Besides the group representation $\pi_{G}$, a representation based
on functions $\phi\left(X\right)$ of coordinates on $\g^{*}$ (thus
analogous of standard functions on momentum space) can also be defined:
``the algebra representation''. The underlying non-commutativity
of the Lie algebra has translate into an appropriately defined non-commutativity
of the functions $\phi\left(X\right)$. This is done by the introduction
of a $\star$-product compatible, in the sense of diagram (\ref{productDiagram}),
with the coproduct $\Delta_{G}$ of the algebra of operators $\hat{\z}^{i}$
corresponding to a specific parametrization of $G$. As we will see,
the required $\star$-product can be defined directly from the choice
of quantization map ${\cal Q}$.

More precisely, one takes coordinates $\z:\ G\rightarrow\g\cong\mathbb{R}^{d}$
on $G$ satisfying $\z_{k}^{i}(\vec{0})=0$, $\frac{\partial}{\partial k^{j}}\z_{k}^{i}(\vec{0})=\delta_{j}^{i}$,
where $k$ are canonical coordinates. The coordinates operators $\hat{\z}^{i}$
are defined by imposing $\hat{f}\stackrel{!}{=}f_{\z}\left(\hat{\z}^{i}\right)$,
where $f_{\z}\circ\vec{\z}\equiv f$ for all $f\in C^{\infty}\left(G\right)$.
The algebra representation $\pi_{\g*}$ then is defined by the actions
\begin{equation}
\begin{gathered}\left(\pi_{\g^{*}}\left(\hat{X}_{i}\right)\phi\right)\left(X\right)\equiv X_{i}\star\phi\left(X\right),\\
\left(\pi_{\g^{*}}\left(\hat{\z}^{i}\right)\phi\right)\left(X\right)\equiv-i\partial^{i}\phi\left(X\right),
\end{gathered}
\label{algebraRep}
\end{equation}
where the second equation means that 
\begin{equation}
\left(\pi_{\g^{*}}\left(\hat{f}\right)\phi\right)\left(X\right)=f_{k}\left(-i\vec{\partial}\right)\phi\left(X\right),
\end{equation}
where $f_{k}(k):=\ f(e^{ik})\in C^{\infty}(\g)$ for all $f\in C^{\infty}(G)$.
The algebra representation is thus defined on the function space $L_{\star}^{2}\left(\g^{*}\right)\ni\phi\left(X\right)$,
which denotes the quotient $C_{c}^{\infty}\left(\g^{*}\right)/{\cal N}$
of the space of smooth compactly supported functions (see footnote
\ref{fn:compactness}) on $\g^{*}$ by the subspace ${\cal N}$, such
that the inner product respect to the Lebesgue measure $d^{d}X$ on
$\g^{*}$ 
\begin{equation}
\left\langle \phi,\phi'\right\rangle _{\g^{*}}:=\int_{\g^{*}}\frac{d^{d}X}{\left(2\pi\right)^{d}}\left(\bar{\phi}\star\phi'\right)\left(X\right)\label{L2Xnorm}
\end{equation}
is non degenerate, i.e. with ${\cal N}:=\left\{ \phi\in C^{\infty}\left(\g^{*}\right):\left\langle \phi,\phi'\right\rangle _{\g^{*}}=0\right\} $.
The first equation in (\ref{algebraRep}) is actually a consequence
of a stronger condition that one imposes on the $\star$-product:
\begin{equation}
(\pi_{\g^{*}}(f(\hat{X}_{i}))\phi\left(X\right))\equiv f_{\star}\left(X\right)\star\phi\left(X\right),
\end{equation}
for all $f_{\star}\in{\cal A}_{\g^{*}}\subset C^{\infty}\left(\g^{*}\right)$
such that $f(\hat{X}_{i})={\cal Q}\left(f_{\star}\right)\in\U_{\g^{*}}$.
With this definition the $\star$-product and quantization map are
related, as in the deformation quantization framework, by 
\begin{equation}
f_{\star}\star f_{\star}'={\cal Q}^{-1}({\cal Q}(f_{\star}){\cal Q}(f_{\star}')).\label{starQrelation}
\end{equation}
As anticipated, the choice of quantization map determines uniquely
the $\star$-product to be used in constructing the algebra representation.

The faithfulness of the representation (\ref{algebraRep}) is then
guaranteed by noticing that the $\star$-product is such that the
observables depending on $\hat{X}_{i}$ are represented through an
algebra isomorphism, while $f\rightarrow\pi_{\g^{*}}(\hat{f})$ is
a homomorphism due to commutativity of $\partial^{i}$ on $\g^{*}$.
Finally, one can show (see \cite{GuOrRaNCFT}) that the commutator
$[\hat{X}_{i},\hat{\z}^{j}]=i\widehat{{\cal L}_{i}\z^{j}}$ is reproduced
if the $\star$-product satisfies the property encoded in the commutative
diagram (\ref{productDiagram}), i.e. if 
\begin{equation}
\pi_{\g^{*}}(\hat{\z}^{i})\circ m_{\g^{*}}=m_{\g^{*}}\circ(\pi_{\g^{*}}\otimes\pi_{\g^{*}})(\Delta_{G}(\hat{\z}^{i})).\label{compatibilityAlgebra}
\end{equation}
Thus, the compatibility condition (\ref{productDiagram}) can also
be understood as expressing the compatibility between the choice of
quantization map, i.e. of $\star$-product, and the choice of coordinates
on the group used in defining the quantum algebra $\U$. For more
details on the algebra representations, see \cite{GuOrRaNCFT}.

\subsection{The non commutative Fourier transform}

The non-commutative Fourier transform ${\cal F}:\ L^{2}\left(G\right)\rightarrow L_{\star}^{2}\left(g^{*}\right)$
(and its inverse ${\cal F}^{-1}:\ L_{\star}^{2}\left(g^{*}\right)\rightarrow L^{2}\left(G\right)$)
is the intertwiner between the group and algebra representation (and
viceversa), ensuring their unitary equivalence.

It is expressed as the integral transform: 
\begin{equation}
\begin{gathered}\tilde{\psi}\left(X\right):={\cal F}\left(\psi\right)\left(X\right)=\int_{G}dg\ E_{g}\left(X\right)\psi\left(g\right)\in L_{\star}^{2}\left(\g^{*}\right),\\
\psi\left(g\right):={\cal F}^{-1}\left(\tilde{\psi}\right)\left(g\right)=\int_{\g^{*}}\frac{d^{d}X}{\left(2\pi\right)^{d}}\ \overline{E_{g}\left(X\right)}\star\tilde{\psi}\left(X\right)\in L^{2}\left(G\right).
\end{gathered}
\label{FourierTransDef}
\end{equation}
The kernel of the transformation $E_{g}\left(X\right)$ is called
the non-commutative plane wave. Its defining equations are identified
by requiring that the intertwined function spaces define a representation
of the same quantum algebra, and applying the action of $\U$ in the
different representation. The explicit form of $E_{g}\left(X\right)$,
and its existence, depends then again on the choice of a quantization
map, and thus, in turn, on the choice of a deformation quantization
$\star$-product.

The intertwining property can be expressed generally as ${\cal F}\circ\pi_{G}(\hat{T})=\pi_{\g^{*}}(\hat{T})\circ{\cal F}$,
where $\hat{T}\in\U$. Applying it to the operators $\hat{X}_{i}$
and $\hat{\zeta}^{i}$ respectively, one obtains the following conditions
to be satisfied by the Kernel $E_{g}\left(X\right)$: 
\begin{equation}
\begin{gathered}-i{\cal L}_{i}E_{g}\left(X\right)=X_{i}\star E_{g}\left(X\right),\\
-i\partial^{i}E_{g}\left(X\right)=\z^{i}\left(g\right)E_{g}\left(X\right),
\end{gathered}
\label{defEgx}
\end{equation}
which can be integrated (modulo the non-trivial global properties
of $G$) as 
\begin{equation}
\begin{gathered}E\left(hg,X\right)=e^{k(h)\cdot\vec{{\cal L}}}E(g,X)=e_{\star}^{ik\left(h\right)\cdot X}\star E\left(g,X\right),\\
E\left(g,X+Y\right)=e^{Y\cdot\vec{\partial}}E\left(g,X\right)=e^{i\zeta\left(g\right)\cdot Y}E\left(g,X\right),
\end{gathered}
\label{defIntEgX}
\end{equation}
where $k\left(h\right)=-i\ln\left(h\right)\in\g$ are canonical coordinates,
and 
\begin{equation}
e_{\star}^{f\left(X\right)}=\sum_{n=0}^{\infty}\frac{1}{n!}\underbrace{f\star\cdots\star f}_{n\ \textrm{times}}\left(X\right)
\end{equation}
Combining the two expressions one can show that the plane wave is
given by the $\star$-exponential in terms of the canonical coordinates
on the group manifold 
\begin{equation}
E_{g}\left(X\right)=e_{\star}^{ik\left(g\right)\cdot X}=\sum_{n=0}^{\infty}\frac{i^{n}}{n!}k^{j_{1}}\left(g\right)\cdots k^{j_{n}}\left(g\right)X_{j_{1}}\star\cdots\star X_{j_{n}}.\label{starWaveExpansion}
\end{equation}
The integral form of the plane wave is subject to the possible non-trivial
global properties of $G$. For the definition of the Fourier transform,
since $E\left(g,X\right)$ has to be considered only under integration,
weak exponentiality of $G$ is a sufficient condition. Moreover, in
case $G$ has compact subgroups, the logarithm map is multivalued,
and the definition in terms of canonical coordinates has to refer
only to one of the charts covering the group manifold, and extended
appropriately beyond it (this can be done in several ways, see \cite{GuOrRaNCFT}
and references cited there). Let us discuss this point a little further,
briefly.

From the relation (\ref{starQrelation}) it follows that the product
of plane waves is given by 
\begin{equation}
E_{g}\left(X\right)\star E_{h}\left(X\right)={\cal Q}^{-1}(e^{ik\left(g\right)\cdot\hat{X}}e^{ik\left(h\right)\cdot\hat{X}})=e_{\star}^{i{\cal B}\left(k\left(g\right),k\left(h\right)\right)\cdot X},\label{eq:productE}
\end{equation}
where ${\cal B}\left(k\left(g\right),k\left(h\right)\right)$ is given
by the application of the Baker-Campbell-Hausdorff (BCH) formula to
$k\left(g\right),k\left(h\right)\in\g$, and 
\begin{equation}
{\cal Q}\left(e_{\star}^{ik\cdot X}\right)=e^{ik\cdot\hat{X}}.\label{StarExponential}
\end{equation}
In order to have a one-to-one relation between elements of the set
of plane waves ${\cal E}:=\{e^{ik.\hat{X}}:k\in\g\}\subset\U_{\g^{*}}$
and the group $G$, one can define the coordinates $k\left(g\right)=-i\ln\left(g\right)$
to lie in the principal branch, and amend the $\star$-product by
a projection onto the principal branch of the logarithm, $\star\rightarrow\star_{p}$.
With this definition the set of plane waves constitutes a representation
of $G$ respect to the $\star_{p}$-product, since\footnote{Without the projection on the principal branch, the coordinate ${\cal B}^{i}\left(k\left(g\right),k\left(h\right)\right)$
in (\ref{eq:productE}) may lie in any branch of the logarithm.} 
\begin{equation}
E_{g}\left(X\right)\star_{p}E_{h}\left(X\right)=E_{gh}\left(X\right).\label{EgstarEh}
\end{equation}
We will omit in the following the suffix $p$ and implicitly assume
that the projection on the principal branch is implemented.

Thus, in terms of canonical coordinates $k\left(g\right)$ on the
group, the non-commutative plane waves takes the form of a $\star$-exponential
with respect to the $\star$-product corresponding to the chosen quantization
map.

However, from the second of Eqs. (\ref{defIntEgX}), it follows also
that there exists a choice of coordinates $\z^{i}\left(g\right)$
such that the same star-exponentials take the form of classical exponentials
\begin{equation}
E_{g}\left(X\right)={\cal A}\left(g\right)e^{i\z\left(g\right)\cdot X}.\label{planeWaveZeta}
\end{equation}
with a multiplicative prefactor ${\cal A}\left(g\right)=E_{g}\left(0\right)$.
The choice of such coordinates $\z^{i}\left(g\right)$ and the corresponding
prefactor ${\cal A}\left(g\right)$ are also determined by the quantization
map. For this second form the multivaluedness issue affects the choice
of coordinates $\z\ :\ G\rightarrow\g$.

Finally, it is possible to show (see \cite{GuOrRaNCFT}) that when
the $\star$-product is verified to lead to a non-commutative plane
wave of the form (\ref{planeWaveZeta}), i.e. when $E_{g}\left(X\right)=e_{\star}^{ik\left(g\right)\cdot X}={\cal A}\left(g\right)e^{i\z\left(g\right).X}$,
then the $\z$-coordinates satisfy the compatibility equation (\ref{compatibilityAlgebra}),
i.e. they are compatible with the $\star$-product in the sense of
the commutative diagram (\ref{productDiagram}). This means that the
algebra representation is guaranteed to exist, and to be obtained
by non-commutative Fourier transform from the group representation,
only if such coordinates can be found.

\subsection{Summary}

Let us summarize schematically the main elements introduced so far.
One starts from the phase space of a system whose configuration space
is a Lie group $G$, given by the cotangent bundle $T^{*}G\cong G\times\g^{*}$,
and its Poisson algebra ${\cal P}_{G}$. A quantization of (a maximal
subalgebra ${\cal A}$ of) ${\cal P}_{G}$ can be obtained through
a quantization map ${\cal Q}$, leading to the quantum algebra $\U\approx\U_{G}\times\U_{\g^{*}}$.
The latter is endowed, for each of its subspaces $\U_{G}$ and $\U_{\g^{*}}$,
with a natural Hopf-algebra structure characterized by $\left(\Delta_{G},\epsilon_{G},S_{G}\right)$
and $\left(\Delta_{\g^{*}},\epsilon_{\g^{*}},S_{\g^{*}}\right)$.

Two different representation of $\U$ can be defined, the group representation
$\pi_{G}$ on $L^{2}\left(G\right)$, and the algebra representation
$\pi_{\g^{*}}$on $L_{\star}^{2}\left(\g^{*}\right)$. The latter
is characterized by the definition of a suitable $\star$-product
compatible with the quantization map, and determined by it. Both representations
satisfy the compatibility condition between their natives product
and coproduct, as depicted in the diagram (\ref{productDiagram}),
which can then be understood as a condition for the existence of the
representation.

The non-commutative Fourier transform ${\cal F}:\ L^{2}\left(G\right)\rightarrow L_{\star}^{2}\left(g^{*}\right)$
is the intertwiner between the group and algebra representation, ensuring
their unitary equivalence. Its kernel is the non-commutative plane
wave $E_{g}\left(X\right)$, whose existence and precise expression
depend again on the chosen quantization map, and thus of $\star$-product.
The explicit form of the plane wave can indeed be written in terms
of $\star$-exponentials $E_{g}\left(X\right)=\exp_{\star}\left(k\left(g\right)\cdot X\right)$,
where $k\left(g\right)=-i\ln\left(g\right)$ are canonical coordinates,
or as ordinary exponentials (with a prefactor) $E_{g}\left(X\right)={\cal A}\left(\z\left(g\right)\right)\exp\left(\z\left(g\right)\cdot X\right)$
in terms of coordinates $\z$ such that $\z_{k}\left(0\right)=0$
and $\frac{\partial}{\partial k^{i}}\z_{k}^{j}\left(0\right)=\delta_{i}^{j}$
($\z_{k}\left(k\right)=\z\left(e^{ik}\right)$). The existence of
these kind of coordinates ensures that the compatibility condition
(\ref{productDiagram}) is fulfilled, tying together the existence
of the non-commutative Fourier transform with the one of the algebra
representation.


\section{The Duflo map and Lie algebra representation for the Lorentz group}

\label{sec:dufloLorentz-algebra}

We will now proceed to constructing the structures introduced in the
previous section for the case in which the configuration space of
our system is the Lorentz group. We will first look for suitable coordinates
in which to define the non-commutative plane waves. As explained,
the definition of the plane wave depends on the quantization map,
which in turn specifies the $\star$-product structure. We will work
the Duflo quantization map~\cite{duflo,kontsevichDuflo,CalaqueRossiDuflo}
, which has already found a number of applications in the quantum
gravity context~\cite{AlekseevDuflo,FreMajDuflo,SahlThiemDuflo,SahlThiemDufloPRL,NouiPerezPranzDuflo,VitaleDuflo}.
After defining the properties of the Duflo map, we will derive the
explicit expression of the Duflo plane wave for the Lorentz group,
as a star exponential, and as the ordinary exponential with prefactor
in terms of suitable coordinates on the group. We will then discuss
the properties of the algebra representation and show explicitly the
structure of the star-product for the lowest order polynomials.

These structures, in particular the plane waves, will finally allow
us to define the non-commutative Fourier transform for the Lorentz
group for the Duflo quantization map, in the following section \ref{sec:NCfourier}.

\subsection{Parametrization of the Lorentz group}

\label{sec:SO31}

As discussed, the definition of the non-commutative plane wave, and
thus of the non-commutative Fourier transform, requires the group
to be at least weakly exponential. Exponentiability indeed ensures
that the image of the exponential map $\exp:\ \g\rightarrow G$ is
onto, i.e. that all the points of $G$ are obtained through exponentiation
of its Lie algebra. Also, if a group is weakly exponential the image
of $\exp:\ \g\rightarrow G$ is dense in $G$. Since the non-commutative
Fourier transform is defined only under integration, weak exponentiability
is a sufficient condition for its uniqueness, as the canonical coordinates
$k\left(g\right)=-i\ln\left(g\right)$ defining the plane waves will
cover the whole group except for a set of points of zero measure.
So, in order to define the non-commutative Fourier transform for the
Lorentz group we must first discuss the relation between its parametrization
and the exponential map (we follow mainly the discussion in~\cite{ruhl}).

The (homogeneous) Lorentz group is usually defined by two basic matrix
representations: as the group\footnote{We restrict in this manuscript to the so-called proper orthocronous
Lorentz group, the identity component of O(3,1), consisting of the
set of Lorentz transformations that preserve the orientation of ``spatial\textquotedblright{}
and ``temporal\textquotedblright{} dimensions. } SO(3,1) of real orthogonal 4x4 matrices with unit determinant, preserving
the bilinear form of Lorentzian signature (3,1), or as the group SL(2,$\C$)
of complex unimodular 2x2 matrices. As topological groups, SL(2,$\C$)
is the double covering of SO(3,1). Moreover since SL(2,$\C$) is simply
connected it is also the universal covering of SO(3,1), while SO(3,1)
is isomorphic to SL(2,$\C$)/$\left\{ I,-I\right\} $, where $I$
and $-I$ are the central elements of SL(2,$\C$), and thus SO(3,1)
is doubly-connected.

While for a compact connected (matrix) group one can prove that the
exponential map is surjective~\cite{hall}, when the group is non-compact,
as for SO(3,1) or SL(2,$\C$), surjectiveness is not guaranteed. In
particular one can show that SO(3,1) is exponential, while SL(2,$\C$)
is not~\cite{rossmann}, i.e. not all the points of SL(2,$\C$) are
in the image of the expoential map applied to its Lie algebra $\sl$(2,$\C$).
However, a complex connected Lie group, as is the case of SL(2,$\C$),
is always weakly exponential~(see for instance \cite{hofmanExp}),
so that the image of $\exp:\ \sl\left(2,\C\right)\rightarrow\text{SL}(2,\C)$
is dense. As we stressed, this is all that is needed for a well-defined
non-commutative Fourier transform and algebra representation.

An element of SL(2,$\C$) can be parametrized as 
\begin{equation}
a=a_{0}\id_{2}+\vec{a}\cdot\vec{\sigma},\label{SL(2,C)element}
\end{equation}
where $\id_{2}$ is the identity matrix in 2 dimensions, $a_{0},a_{j},$
with $j=1,2,3,$ are complex numbers, and $\sigma_{j}$ are Pauli
matrices (hermitian traceless 2$\times$2 matrices) 
\begin{equation}
\sigma_{1}=\left(\begin{array}{cc}
0 & 1\\
1 & 0
\end{array}\right),\qquad\sigma_{2}=\left(\begin{array}{cc}
0 & -i\\
i & 0
\end{array}\right),\qquad\sigma_{3}=\left(\begin{array}{cc}
1 & 0\\
0 & -1
\end{array}\right),
\end{equation}
satisfying the Lie brackets 
\begin{equation}
\left[\sigma_{i},\sigma_{j}\right]=2i\epsilon_{ijk}\sigma_{k}.
\end{equation}
The conjugacy classes of SL(2,$\C$) are instead represented by the
matrices~\cite{rossmann,hofmanExp} 
\begin{equation}
\left(\begin{array}{cc}
\alpha & 0\\
0 & \alpha^{-1}
\end{array}\right),\qquad\left(\begin{array}{cc}
1 & 1\\
0 & 1
\end{array}\right),\qquad\left(\begin{array}{cc}
-1 & 1\\
0 & -1
\end{array}\right).\label{conjugClassesSL(2,C)}
\end{equation}

Canonical coordinates (of the first kind) on SL(2,$\C$) are defined~\cite{hall,rossmann}
by applying the exponential map to its Lie algebra $\sl\left(2,C\right)$.
One way of realizing this is as the complexification of SU(2) by means
of a complex rotation vector~\cite{ruhl} 
\begin{equation}
\zeta_{i}=\rho_{i}+i\beta_{i},\qquad i=1,2,3\label{complexRotation}
\end{equation}
where $\rho_{i},\beta_{i}$ are real parameters, so that, choosing
the basis of $\sl\left(2,C\right)$ represented by Pauli matrices,
an element of SL(2,$\C$) is written as 
\begin{equation}
a=\exp\left(\frac{i}{2}\vec{\zeta}\cdot\vec{\sigma}\right).\label{SL(2,C)complexVector}
\end{equation}
With this definition one gets the relations 
\begin{equation}
\begin{gathered}a_{0}=\cos\left(\frac{1}{2}\left(\phi+i\eta\right)\right)\hspace{1cm}a_{j}=i\frac{\sin\left(\frac{1}{2}\left(\phi+i\eta\right)\right)}{\phi+i\eta}\zeta_{j},\end{gathered}
\label{SL(2,C)complexRotation}
\end{equation}
where we define the complex rotation angle $\phi+i\eta$, with $\eta\geq0$,
related to $\vec{\zeta}$ by (denoting $x^{2}=\vec{x}\cdot\vec{x}$)
\begin{equation}
\begin{gathered}\left(\phi+i\eta\right)^{2}=\zeta^{2},\\
\phi^{2}-\eta^{2}=\rho^{2}-\beta^{2},\qquad\phi\eta=\vec{\rho}\cdot\vec{\beta}.
\end{gathered}
\label{phietarhobeta}
\end{equation}

The meaning of $\phi$ and $\eta$ can be better understood considering
the action of a SL(2,$\C$) matrix (\ref{SL(2,C)element}) on a point
in Minkowski spacetime $x=\left(x^{0},x^{1},x^{2},x^{3}\right)$ defined
as 
\begin{equation}
\boldsymbol{x}=x^{0}\id+\vec{x}\cdot\vec{\sigma}=\left(\begin{array}{cc}
x^{0}+x^{3} & x^{1}-ix^{2}\\
x^{1}+ix^{2} & x^{0}-x^{3}
\end{array}\right),
\end{equation}
so that 
\begin{equation}
\boldsymbol{x}'=a\boldsymbol{x}a^{\dagger}\rightarrow x'=\Lambda\left(a\right)x.\label{Lambda(a)}
\end{equation}

The transformation matrices $\Lambda\left(a\right)$ form the elements
of the group SO(3,1). This is identical to the group of linear transformations
leaving invariant the metric $\eta=\text{diag}\left\{ 1,-1,-1,-1\right\} $,
described by $4\times4$ unimodular orthogonal matrices generated
by the set of (purely imaginary traceless) matrices defined by 
\begin{equation}
\left(M_{ab}\right)_{kl}=i\left(\eta_{ak}\delta_{bl}-\eta_{al}\delta_{bk}\right),\qquad a,b,k,l=0,1,2,3,4
\end{equation}
with commutation rules 
\begin{equation}
\left[M_{ab},M_{cd}\right]=i\left(\eta_{ad}M_{bc}+\eta_{bc}M_{ad}-\eta_{ac}M_{bd}-\eta_{bd}M_{ac}\right),
\end{equation}
so that an element of SO(3,1) is (repeated indexes are summed, throughout
the manuscript) 
\begin{equation}
\Lambda\left(\alpha\right)=\exp\left(\frac{i}{2}\alpha_{ij}M_{ij}\right).
\end{equation}
Defining the generators 
\begin{equation}
J_{i}=\frac{1}{2}\epsilon_{ijk}M_{jk},\qquad N_{i}=M_{0i},\qquad i,j=1,2,3,\label{SO31generators}
\end{equation}
corresponding to rotation and boost transformations, satisfying the
commutation rules 
\begin{equation}
\left[J_{i},J_{j}\right]=i\epsilon_{ijk}J_{k},\qquad\left[J_{i},N_{j}\right]=i\epsilon_{ijk}N_{k},\qquad\left[N_{i},N_{j}\right]=-i\epsilon_{ijk}J_{k},\label{eq:LieAlgebraSO31}
\end{equation}
we can express an element of the group as 
\begin{equation}
\Lambda\left(\vec{\rho},\vec{\beta}\right)=\exp\left\{ i\vec{\rho}\cdot\vec{J}+i\vec{\beta}\cdot\vec{N}\right\} ,\label{Lambda(J,N)}
\end{equation}
where 
\begin{equation}
\rho_{i}=\frac{1}{2}\epsilon_{ijk}\alpha_{jk},\qquad\beta_{i}=\alpha_{0i}.
\end{equation}
The explicit expression of $\Lambda\left(\vec{\rho},\vec{\beta}\right)$
is given in the appendix \ref{sec:SO31explicit}, where the relation
between SO(3,1) and SL(2,$\C$) is also discussed. It is important
to notice that the parameters domain is different for the two groups:
the multivaluedness of the logarithmic map (the inverse of the exponential
map) results from the periodicity of the compact subgroup of rotations.
As shown in App. \ref{sec:SO31explicit}, analogously to the relation
between SO(3) and SU(2), SL(2,$\C$) ``covers twice'' SO(3,1), with
the isomorphism $\text{SO}(3,1)\simeq\text{SL(2,}\C)/\left\{ \id,-\id\right\} $.

In order to show the weak exponentiability of the Lorentz group, one
can study the behaviour of the logarithm map by inverting relation
(\ref{SL(2,C)complexVector}). The study of the branch points for
the logarithmic map for the Lorentz group is performed in App. \ref{sec:branchCuts}.
Restricting it to its principal values, the complex rotation vector
is holomorphic in $a_{0}$ except for a branch cut extending on the
real axis of $a_{0}$ from -1 to $-\infty$. Thus, except for the
branch cut, the canonical coordinates provided by the exponential
map, represented by the complex rotation vector $\vec{\zeta}=\vec{\rho}+i\vec{\beta}$,
are holomorphic functions parametrizing the whole SL(2,$\C$) group.
Accordingly, the corresponding range of values of $\phi$ for which
the complex rotation vector is single-valued is $\phi\in(-2\pi,2\pi]$
for $\text{SL(2,}\C)$ and $\phi\in(-\pi,\pi]$ for $\text{SO}(3,1)$.
This defines the principal branch respectively for the $\text{SL(2,}\C)$
and $\text{SO}(3,1)$ parametrizations.

Thus, in the following, we consider a generic element of the Lorentz
group in its exponential representation (\ref{Lambda(J,N)}) or (\ref{SL(2,C)complexVector})
\begin{equation}
g=\exp\left\{ i\vec{\rho}\cdot\vec{J}+i\vec{\beta}\cdot\vec{N}\right\} ,\label{gEXP}
\end{equation}
where the generators $J_{i}$ and $N_{i}$ are the ones defined in
this section, and, depending on the chosen group representation, can
be considered to be the generators (\ref{SO31generators}) of SO(3,1),
$J_{i}=\frac{1}{2}\epsilon_{ijk}M_{jk}$ and $N_{i}=M_{0i}$, or the
generators $J_{i}\equiv\frac{1}{2}\sigma_{i}$ and $N_{i}\equiv\frac{1}{2}i\sigma_{i}$
of the real\footnote{$\sl(2,\C)$ can be considered a real Lie algebra generated of dimension
6 with basis vectors $\left\{ \sigma_{i}\right\} ,\left\{ i\sigma_{i}\right\} $,
not to be confused with its real forms $\su$(2) or $\sl$(2,$\R$).} Lie algebra $\sl(2,\C)_{R}$, satisfying in both cases the Lie brackets
(\ref{eq:LieAlgebraSO31}), the distinction between the two groups
being given by the domain of the ``rotation'' parameter $\phi$
as discussed above. An element of the Lie algebra is then $x=x^{i}e_{i}\in\g$,
with $e_{i}$ the basis of generators $e_{i}\equiv\left(J_{i},N_{i}\right)$,
and $x^{i}$ the set of canonical coordinates $k^{i}\equiv\left(\rho^{i},\beta^{i}\right)$.

\subsection{The Duflo quantization map for the Lorentz group and the non-commutative
plane waves\label{sec:Duflo}}

The Duflo map was introduced in~\cite{duflo} as a generalization
to arbitrary finite-dimensional Lie algebras of the Harish-Chandra
isomorphism between invariant polynomials on the dual of a Lie algebra
and the center of the corresponding universal enveloping algebra.
The isomorphism was proved in a different form within the formalism
of deformation quantization of general Poisson manifolds~\cite{kontsevichDuflo}
(see also~\cite{CalaqueRossiDuflo}), and used then in different
contexts in the quantum gravity literature~\cite{AlekseevDuflo,FreMajDuflo,SahlThiemDuflo,SahlThiemDufloPRL,NouiPerezPranzDuflo,VitaleDuflo}.
An important property of the Duflo map is that it realizes the isomorphism
between the set Sym$(\g)^{\g}$ of invariant polynomials under the
(adjoint) action of $G$, and the set $U\left(\g\right)^{\g}$ of
$G$-invariant differential operators on the enveloping algebra, corresponding
to the center of the enveloping algebra, $U\left(\g\right)^{\g}={\cal Z}\left(U\left(\g\right)\right)$.
Thus, invariant polynomials Sym$(\g)^{\g}$ map to Casimirs of the
quantum algebra, and the map preserves the algebra structure of such
polynomials, so the Duflo quantization map ${\cal D}$ has the remarkable
property that for two elements $\alpha$ and $\beta$ of $U\left(\g\right)^{\g}$,
${\cal D}\left(\alpha\right){\cal D}\left(\beta\right)={\cal D}\left(\alpha\beta\right)$.

Concretely, the Duflo map is defined by the composition 
\begin{equation}
{\cal D}={\cal S}\circ j^{\frac{1}{2}}\left(\partial\right)\label{DufloDef}
\end{equation}
of the ``symmetrization map'' $S$, 
\begin{equation}
{\cal S}\left(X_{i_{1}}\cdots X_{i_{n}}\right)=\frac{1}{n!}\sum_{\sigma\in S_{n}}\hat{X}_{i_{\sigma_{1}}}\cdots\hat{X}_{i_{\sigma_{n}}},\label{symmMap}
\end{equation}
with $S_{n}$ is the symmetric group of order $n$, with the function
on the Lie algebra $\g$ 
\begin{equation}
j\left(x\right)=\det\left(\frac{\sinh\frac{1}{2}\text{ad}_{x}}{\frac{1}{2}\text{ad}_{x}}\right),\qquad x\in\g,\label{j(X)}
\end{equation}
where in (\ref{DufloDef}) one substitutes $x^{i}$ with $\partial^{i}=\partial/\partial X_{i}$
where $X\in\g^{*}$. Here $\text{ad}_{x}$ is in the adjoint representation
of $\g$. Notice also that with this definition the Duflo function
(\ref{j(X)}) coincides with the Jacobian of the exponential map\textbf{~}\cite{AlekDufloJacobian,DufloJacobian2007},
i.e. for $g=\exp\left(x\right)$ 
\begin{equation}
dg=j\left(x\right)dx.\label{DufloJacobian}
\end{equation}

As we have explained in Sec. \ref{sec:definitions}, once we have
chosen the Duflo quantization map, the compatibility between the associated
$\star$-product, given by relation (\ref{starQrelation}) as 
\begin{equation}
f_{\star}\star f_{\star}'={\cal D}^{-1}({\cal D}(f_{\star}){\cal D}(f_{\star}')),\label{DufloStarQ}
\end{equation}
and the coproduct on the algebra of functions on the group, and thus
the existence of the algebra representation, is guaranteed by the
construction of the non-commutative plane wave, the kernel of the
transformation between the group and algebra representations. Our
next goal is then to derive the explicit expression of this non-commutative
plane wave, and in particular its expression as a standard exponential
(with a certain prefactor) in terms of specific coordinates on the
group.

We need first to calculate the function $j\left(x\right)$ for the
Lorentz group. We consider a generic element of the Lorentz group
written in its exponential representation (\ref{Lambda(J,N)}) or
(\ref{SL(2,C)complexVector}), 
\begin{equation}
g=\exp\left\{ i\vec{\rho}\cdot\vec{J}+i\vec{\beta}\cdot\vec{N}\right\} ,
\end{equation}
where the generators $J_{i}$ and $N_{i}$ are the ones defined in
the previous section, and can be considered to correspond to the $J_{i}$,$N_{i}$
matrices generating SO(3,1) or to the generators $J_{i}\equiv\frac{1}{2}\sigma_{i}$,
$N_{i}\equiv\frac{1}{2}i\sigma_{i}$ of the real\footnote{$\sl(2,\C)$ can be considered to be a real Lie algebra generated
of dimension 6 with basis vectors $\left\{ \sigma_{i}\right\} ,\left\{ i\sigma_{i}\right\} $,
not to be confused with one of its real forms $\su$(2) or $\sl$(2,$\R$).} Lie algebra $\sl(2,\C)_{R}$, in both cases satisfying the Lie brackets
(\ref{eq:LieAlgebraSO31}), the distinction between the two groups
being given by the domain of the rotation parameter $\phi$ as discussed
in the previous section. The calculation of the the Duflo function
$j\left(x\right)$ on the element (\ref{gEXP}) is performed in App.
\ref{sec:DufloFactor}, and, for an element of the Lie algebra $x=x_{J}^{i}J_{i}+x_{N}^{i}N_{i}$,
gives 
\begin{equation}
j^{\frac{1}{2}}\left(x\right)=4\frac{\left|\sinh\left(\tfrac{1}{2}x_{\zeta}\right)\right|^{2}}{\left|x_{\zeta}^{2}\right|},\label{eq:DufloFunction}
\end{equation}
where $x_{\zeta}=\sqrt{\vec{x}_{\zeta}\cdot\vec{x}_{\zeta}}$ and
$x_{\zeta}^{i}=x_{J}^{i}+ix_{N}^{i}$. We can now apply the Duflo
the function $j^{\frac{1}{2}}\left(\partial\right)$ on exponentials
$\exp\left(ik^{i}X_{i}\right)=\exp\left(i\rho^{i}X_{i}^{J}+i\beta^{i}X_{i}^{N}\right)$,
with coordinates $k^{i}\equiv\left(\rho^{i},\beta^{i}\right)$ corresponding
to canonical coordinates on the group $k\left(g\right)=-i\ln\left(g\right)$,
and coordinates on $\g^{*}$ $X_{i}\equiv\left(X_{i}^{J},X_{i}^{N}\right)$
associated respectively to the $J_{i}$ and $N_{i}$ generators\footnote{More precisely $X_{i}$ are the coordinates on the Lie algebra $\g^{*}$
defined by the basis $\left\{ \tilde{e}_{i}\right\} $ dual to $\left\{ e_{i}\right\} \equiv\left\{ J_{i},N_{i}\right\} $
with the canonical pairing $\left\langle \tilde{e}_{i},e_{j}\right\rangle =\delta_{ij}$,
$X=X_{i}\tilde{e}_{i}\in\g^{*}$.}. Thus, after the substitution $x_{J}^{i}\rightarrow\partial_{J}^{i},\ x_{N}^{i}\rightarrow\partial_{N}^{i}$
(i.e. $x_{\zeta}^{i}\rightarrow\partial_{\zeta}^{i}=\partial_{J}^{i}+i\partial_{N}^{i}$),
we obtain 
\begin{equation}
\begin{split} & \left(j^{\frac{1}{2}}\left(\partial\right)\exp\right)\left(i\vec{\rho}\cdot\vec{X}^{J}+i\vec{\beta}\cdot\vec{X}^{N}\right)\\
= & 4\frac{\left|\sin\left(\tfrac{1}{2}\zeta\right)\right|^{2}}{\left|\zeta^{2}\right|}e^{i\vec{\rho}\cdot\vec{X}^{J}+i\vec{\beta}\cdot\vec{X}^{N}}\\
= & 4\frac{\cosh^{2}\left(\tfrac{1}{2}\eta\right)\sin^{2}\left(\tfrac{1}{2}\phi\right)+\sinh^{2}\left(\tfrac{1}{2}\eta\right)\cos^{2}\left(\tfrac{1}{2}\phi\right)}{\phi^{2}+\eta^{2}}e^{i\vec{\rho}\cdot\vec{X}^{J}+i\vec{\beta}\cdot\vec{X}^{N}},
\end{split}
\label{DufloFunctionExp}
\end{equation}
where $\zeta_{i}=\rho_{i}+i\beta_{i}$, and we have considered also
the parametrization (\ref{phietarhobeta}), $\left(\phi+i\eta\right)^{2}=\zeta^{2}$,
$\phi^{2}-\eta^{2}=\rho^{2}-\beta^{2}$, $\phi\eta=\vec{\rho}\cdot\vec{\beta}$.

Notice now that the exponential keeps the same form under the symmetrization
map (\ref{symmMap}): 
\begin{equation}
{\cal S}\left(e^{ik^{i}X_{i}}\right)=e^{ik^{i}\hat{X}_{i}}.
\end{equation}
Using the above results, we finally obtain the action of the Duflo
map on the exponential function 
\begin{equation}
{\cal D}\left(e^{i\vec{\rho}\cdot\vec{X}^{J}+i\vec{\beta}\cdot\vec{X}^{N}}\right)=4\frac{\cosh^{2}\left(\tfrac{1}{2}\eta\right)\sin^{2}\left(\tfrac{1}{2}\phi\right)+\sinh^{2}\left(\tfrac{1}{2}\eta\right)\cos^{2}\left(\tfrac{1}{2}\phi\right)}{\phi^{2}+\eta^{2}}e^{i\vec{\rho}\cdot\hat{X}^{J}+i\vec{\beta}\cdot\hat{X}^{N}}.\label{DufloMapExp}
\end{equation}

The last equation allows us to define the non-commutative plane wave
corresponding to the Duflo quantization map for the Lorentz group.
This can be given in its expression in terms of $\star$-exponential
(\ref{starWaveExpansion}), which in turn encodes the property (\ref{DufloStarQ})
(as in (\ref{StarExponential})), and thus defines the $\star$-product
for the Duflo map. We have indeed 
\begin{equation}
{\cal D}\left(E_{g}\left(X\right)\right)={\cal D}\left(e_{\star}^{i\vec{\rho}\left(g\right)\cdot\vec{X}^{J}+i\vec{\beta}\left(g\right)\cdot\vec{X}^{N}}\right)=e^{i\vec{\rho}\left(g\right)\cdot\hat{X}^{J}+i\vec{\beta}\left(g\right)\cdot\hat{X}^{N}}.\label{D(EgX)}
\end{equation}
At the same time the last equation can be inverted, and using Eq.
(\ref{DufloMapExp}) we get the relation between the plane wave as
a $\star$-exponential and as a standard exponential with prefactor
(\ref{planeWaveZeta}): 
\begin{equation}
E_{g}\left(X\right)={\cal D}^{-1}\left(e^{i\vec{\rho}\left(g\right)\cdot\hat{X}^{J}+i\vec{\beta}\left(g\right)\cdot\hat{X}^{N}}\right)=e_{\star}^{i\vec{\rho}\left(g\right)\cdot\vec{X}^{J}+i\vec{\beta}\left(g\right)\cdot\vec{X}^{N}}={\cal A}\left(\vec{\rho}\left(g\right),\vec{\beta}\left(g\right)\right)e^{i\vec{\rho}\left(g\right)\cdot\vec{X}^{J}+i\vec{\beta}\left(g\right)\cdot\vec{X}^{N}},\label{EstarE}
\end{equation}
with 
\begin{equation}
{\cal A}\left(\vec{\rho}\left(g\right),\vec{\beta}\left(g\right)\right)=\frac{\phi\left(g\right)^{2}+\eta\left(g\right)^{2}}{4\left(\cosh^{2}\left(\tfrac{1}{2}\eta\left(g\right)\right)\sin^{2}\left(\tfrac{1}{2}\phi\left(g\right)\right)+\sinh^{2}\left(\tfrac{1}{2}\eta\left(g\right)\right)\cos^{2}\left(\tfrac{1}{2}\phi\left(g\right)\right)\right)},\label{DufloFactorRho}
\end{equation}
where $\left(\phi,\eta\right)\equiv\left(\phi\left(\vec{\rho},\vec{\beta}\right),\eta\left(\vec{\rho},\vec{\beta}\right)\right)$.
Expression (\ref{EstarE}) shows that within the Duflo quantization
map the coordinates for which the non-commutative plane wave takes
the form of the standard exponential are still the canonical coordinates
$\z\left(g\right)=k\left(g\right)=-i\ln\left(g\right)\equiv\left(\vec{\rho},\vec{\beta}\right)$,
but with a prefactor ${\cal A}\left(\vec{\rho},\vec{\beta}\right)$,
amounting, as expected from Eq. (\ref{DufloJacobian}), to the inverse
of the square root of the Jacobian of the exponential map (see Eq.
\ref{HaarSL2C}). Notice that this is exactly what happens, for example,
in the simpler $SU(2)$ case, studied in ~\cite{GuOrRaNCFT}. We
have thus obtained the explicit form of the plane wave as a standard
exponential, ensuring the existence of the algebra representation,
which we are now going to discuss.

\subsection{The non-commutative algebra representation from the Duflo map}

We mentioned in Sec. \ref{sec:definitions}, and it was proven in
\cite{GuOrRaNCFT}, that the coordinates for which the non-commutative
plane wave can be written as a standard exponential (plus prefactor),
satisfy the compatibility relation (\ref{compatibilityAlgebra}) (or
(\ref{productDiagram})) with the $\star$-product, which, in turn,
ensures the existence of the algebra representation for this choice
of quantization map. In other words, having derived relation (\ref{EstarE}-\ref{DufloFactorRho}),
we have ensured that, for the $\star$-product associated to the Duflo
map, the action (\ref{algebraRep}) on functions $\phi\left(X\right)\in C_{c}^{\infty}\left(\g_{\star}\right)$
($X_{i}\equiv(X_{i}^{J},X_{i}^{N})$ $\z_{i}\equiv\left(\rho_{i},\beta_{i}\right)$,
$\partial^{i}=\partial/\partial X_{i}$) 
\begin{equation}
\begin{gathered}\left(\pi_{\g^{*}}\left(\hat{X}_{i}\right)\phi\right)\left(X\right)\equiv X_{i}\star\phi\left(X\right),\\
\left(\pi_{\g^{*}}\left(\hat{\z}^{i}\right)\phi\right)\left(X\right)\equiv-i\partial^{i}\phi\left(X\right),
\end{gathered}
\label{algebraRep-1}
\end{equation}
is a representation of the quantum algebra (\ref{commOperators})
\begin{equation}
\left[\hat{X}_{i}^{J},\hat{X}_{j}^{J}\right]=i\epsilon_{ij}^{\ k}\hat{X}_{k}^{J},\quad\left[\hat{X}_{i}^{J},\hat{X}_{j}^{N}\right]=i\epsilon_{ij}^{\ k}\hat{X}_{k}^{N},\quad\left[\hat{X}_{i}^{N},\hat{X}_{j}^{N}\right]=-i\epsilon_{ij}^{\ k}\hat{X}_{k}^{N},\label{QalgebraX}
\end{equation}
\begin{equation}
\left[\hat{\rho}_{i},\hat{\rho}_{j}\right]=\left[\hat{\beta}_{i},\hat{\beta}_{j}\right]=\left[\hat{\rho}_{i},\hat{\beta}_{j}\right]=0,\label{Qalgebraz}
\end{equation}
\begin{equation}
\left[\hat{X}_{i},\hat{\z}_{j}\right]=i\widehat{{\cal L}_{X_{i}}\z_{j}}.\label{QalgebraXz}
\end{equation}
While the above statement can be proved in general (see \cite{GuOrRaNCFT})
using the properties of the $\star$-product, we can still show more
explicitly that this is indeed the case for the structures derived
in the previous section. Notice first that relations (\ref{Qalgebraz})
are trivially satisfied by the representation (\ref{algebraRep-1})
due to the commutativity of ordinary derivatives. Let us then evaluate
the $\star$-product between the lowest order monomials. Considering
expressions (\ref{EstarE})\footnote{Obviously, knowing the definition of the Duflo map ~\ref{DufloDef}
and the general relation between the quantization map and the associated
$\star$-product ~\ref{starQrelation}, the action of any operator
in the algebra representation and the explicit expression of any $\star$-monomial
can be computed directly. Having at hand the formula for the exponential
elements, however, allows a more immediate calculation.} the $\star$-product between $n$ coordinates of $\g^{*}$ can be
evaluated through the formula 
\begin{equation}
X_{i_{1}}\star X_{i_{2}}\star\cdots X_{i_{n}}=\left(-i\right)^{n}\frac{\partial^{n}}{\partial k_{1}^{i_{1}}\partial k_{2}^{j_{2}}\cdots\partial k_{3}^{i_{n}}}\Big|_{k_{1}=k_{2}=\cdots=k_{n}=0}{\cal D}^{-1}\left(e^{i\vec{{\cal B}}\left(k_{1},k_{2},\dots,k_{n}\right)\cdot\hat{X}}\right).\label{StarProdCalc}
\end{equation}
From this formula one obtains (the explicit calculation is reported
in App. \ref{sec:starProd}, where third order monomials are also
reported for completeness):

\begin{equation}
\begin{gathered}X_{i}^{J}\star X_{j}^{J}=X_{i}^{J}X_{j}^{J}+\frac{i}{2}\epsilon_{ij}^{\ k}X_{k}^{J}-\frac{1}{6}\delta_{ij},\\
X_{i}^{J}\star X_{j}^{N}=X_{i}^{J}X_{j}^{N}+\frac{i}{2}\epsilon_{ij}^{\ k}X_{k}^{N},\\
X_{i}^{N}\star X_{j}^{N}=X_{i}^{N}X_{j}^{N}-\frac{i}{2}\epsilon_{ij}^{\ k}X_{k}^{J}+\frac{1}{6}\delta_{ij}.
\end{gathered}
\label{StarDouble}
\end{equation}
These relations show immediately that that Eqs. (\ref{QalgebraX})
are fulfilled by the representation (\ref{algebraRep-1}), i.e. that
\begin{equation}
X_{i}\star X_{j}\star\phi\left(X\right)-X_{j}\star X_{i}\star\phi\left(X\right)=i\epsilon_{ij}^{\ k}X_{k}^{J}\star\phi\left(X\right).
\end{equation}

Finally we must check the commutators (\ref{QalgebraXz}). It is enough
to consider the action (\ref{algebraRep-1}) on plane waves (\ref{EstarE}),
since, if the non-commutative Fourier transform can be defined, as
we will do in the next section, any function can be decomposed in
terms of its plane-wave basis. We need to show then that 
\begin{equation}
-iX_{i}\star\partial^{j}E_{g}\left(X\right)+i\partial^{j}\left(X_{i}\star E_{g}\left(X\right)\right)=i\left({\cal L}_{X_{i}}\z_{j}\left(g\right)\right)\Big|_{\z_{j}\left(g\right)=-i\partial^{j}}E_{g}\left(X\right)\label{comXzRep}
\end{equation}
From relations (\ref{defEgx}) defining the plane wave we have that
the l.h.s. can be rewritten as 
\begin{equation}
-i\z_{j}\left(g\right){\cal L}_{X_{i}}E_{g}\left(X\right)+\partial^{j}\left({\cal L}_{X_{i}}E_{g}\left(X\right)\right).\label{comXzRep1}
\end{equation}
The Lie derivative 
\begin{equation}
{\cal L}_{X_{i}}f\left(g\right)=\frac{d}{dt}f\left(e^{ite_{i}}g\right)\Big|_{t=0}
\end{equation}
can be expressed in terms of coordinates $\z^{i}$ for which $E_{g}\left(X\right)={\cal A}\left(\z\left(g\right)\right)\exp\left(i\z\left(g\right)\cdot X\right)$
as 
\begin{equation}
\begin{split}{\cal L}_{X_{i}}E_{g}\left(X\right)=\L_{i}^{j}\left(\z\left(g\right)\right)\partial_{\z_{j}}E_{g}\left(X\right)=\L_{i}^{j}\left(\z\left(g\right)\right)\left(\partial_{\z_{j}}{\cal A}\left(\z\left(g\right)\right)+iX_{j}{\cal A}\left(\z\left(g\right)\right)\right)e^{i\z\left(g\right)\cdot X}\end{split}
\end{equation}
so that 
\begin{equation}
\begin{split}\partial^{j}\left({\cal L}_{X_{i}}E_{g}\left(X\right)\right)=\,i\L_{i}^{j}\left(\z\left(g\right)\right){\cal A}\left(\z\left(g\right)\right)e^{i\z\left(g\right)\cdot X}+\L_{i}^{j}\left(\z\left(g\right)\right)\left(\partial_{\z_{j}}{\cal A}\left(\z\left(g\right)\right)+iX_{j}{\cal A}\left(\z\left(g\right)\right)\right)\partial^{j}e^{i\z\left(g\right)\cdot X}.\end{split}
\end{equation}
We now use the second of relations (\ref{defEgx}), and the relation
${\cal L}_{X_{i}}\z_{j}\left(g\right)=\L_{i}^{j}\left(\z\left(g\right)\right)$
to rewrite the last expression as 
\begin{equation}
i\left({\cal L}_{X_{i}}\z_{j}\left(g\right)\right)E_{g}\left(X\right)+i\z_{j}\left(g\right){\cal L}_{X_{i}}E_{g}\left(X\right),
\end{equation}
which, substituted in (\ref{comXzRep1}), and using again the second
of (\ref{defEgx}), returns (\ref{comXzRep}). Thus, we confirm the
relations (\ref{algebraRep-1}) define a representation of (\ref{QalgebraXz}).
Moreover, as mentioned already, one can show in general (see \cite{GuOrRaNCFT})
that this amounts to the compatibility between (the coproduct for)
the coordinates $\z_{i}\left(g\right)$ and the $\star$-product in
the sense of the commutative diagram (\ref{productDiagram}).


\section{The non-commutative Fourier transform for the Lorentz group}

\label{sec:NCfourier}

The non-commutative plane wave (\ref{EstarE}) is the kernel of the
non-commutative Fourier transform, the transform between the group
and algebra representation. We have now all the material to show the
form of such non-commutative Fourier transform and to discuss some
of its properties. We will discuss also its relation with other known
constructions for a Fourier transform on the group, as well as with
the Plancherel decomposition into irreducible representations.

\subsection{The NC Fourier transform for SL(2,$\C$)}

In order to write the expression of the non-commutative Fourier transform
we first need to evaluate the Haar measure $\mu\left(g\right)$ for
the Lorentz group in canonical coordinates $k^{i}\left(g\right)\equiv\left(\rho^{i}\left(g\right),\beta^{i}\left(g\right)\right)$,
such that $dg=\mu\left(\vec{\rho}\left(g\right),\vec{\beta}\left(g\right)\right)d^{3}\vec{\rho}d^{3}\vec{\beta}$.
The calculation is performed in App. (\ref{sec:HaarSL2C}). From the
definition (\ref{FourierTransDef}), considering the expression of
the plane wave (\ref{EstarE}) and the Haar measure (\ref{HaarSL2C})
the Fourier transform ${\cal F}\ :\ L^{2}\left(g\right)\rightarrow L_{\star}^{2}\left(\g^{*}\right)$
and its inverse ${\cal F}^{-1}$ are 
\begin{equation}
\begin{gathered}\tilde{\psi}\left(X\right)={\cal F}\left(\psi\right)\left(X\right)=\int_{G}d^{3}\vec{\rho}d^{3}\vec{\beta}\ {\cal A}^{-1}\left(\vec{\rho},\vec{\beta}\right)e^{i\vec{\rho}\cdot\vec{X}^{J}+i\vec{\beta}\cdot\vec{X}^{N}}\psi\left(g\right),\\
\psi\left(g\right)={\cal F}^{-1}(\tilde{\psi})\left(g\right)=\frac{1}{\left(2\pi\right)^{6}}\int_{\g^{*}}d^{3}X^{J}d^{3}X^{N}\ {\cal A}\left(\vec{\rho},\vec{\beta}\right)e^{-i\vec{\rho}\cdot\vec{X}^{J}-i\vec{\beta}\cdot\vec{X}^{N}}\star\tilde{\psi}\left(X\right),
\end{gathered}
\label{FourierTransSL2C}
\end{equation}
where $\left(\vec{\rho},\vec{\beta}\right)=\left(\vec{\rho}\left(g\right),\vec{\beta}\left(g\right)\right)$,
and the integration measure $d^{3}\vec{\rho}d^{3}\vec{\beta}$ is
the standard Lebesgue measure, with the parameters restricted as explained
in section (\ref{sec:SO31}) depending if $G$ is $\text{SL}(2,\C)$
or $\text{SO}(3,1)$: $\eta\in[0,\infty)$ and $\phi\in(-2\pi,2\pi]$
for $\text{SL(2,}\C)$ while $\phi\in(-\pi,\pi]$ for $\text{SO}(3,1)$,
with $\phi^{2}-\eta^{2}=\rho^{2}-\beta^{2}$, $\phi\eta=\vec{\rho}\cdot\vec{\beta}$.
The integration measure $d^{3}X^{J}d^{3}X^{N}$ is the Lebesgue measure
$d^{3}X^{J}d^{3}X^{N}$ with $X^{J}\in\R^{3}$, $X^{N}\in\R^{3}$.
As explained in Sec. \ref{sec:definitions}, the Fourier transform
so defined acts as intertwiner between the group and algebra representation,
where the intertwining property can be expressed as ${\cal F}\circ\pi_{G}(\hat{T})=\pi_{\g^{*}}(\hat{T})\circ{\cal F}$,
where $\hat{T}\in\U$.

From the definition of the Fourier transform (\ref{FourierTransSL2C}),
we get the general following properties (we refer to~\cite{GuOrRaNCFT}
for the general proofs): 
\begin{itemize}
\item The group multiplication from the left is dually represented on ${\cal F}\left(\psi\right)\left(X\right)$
as $\star$-multiplication by $E_{g^{-1}}\left(X\right)$: 
\begin{equation}
{\cal F}\left(L_{g}\psi\right)\left(X\right)=\int_{G}dh\ E_{h}\left(X\right)\psi\left(gh\right)=E_{g^{-1}}\left(X\right)\star{\cal F}\left(\psi\right)\left(X\right).
\end{equation}
\item Noticing that $\overline{E_{g}\left(X\right)}=E_{g^{-1}}\left(X\right)=E_{g}\left(-X\right)$
we get the delta function on $L^{2}\left(g\right)$ 
\begin{equation}
\begin{split}\delta\left(g\right)=\delta^{d}\left(\z\left(g\right)\right)= & \int_{\g^{*}}\frac{d^{d}X}{\left(2\pi\right)^{d}}\ \overline{E_{g}\left(X\right)}\\
= & \frac{1}{\left(2\pi\right)^{6}}\int_{\g^{*}}d^{3}X^{J}d^{3}X^{N}\ {\cal A}\left(\vec{\rho},\vec{\beta}\right)e^{-i\vec{\rho}\cdot\vec{X}^{J}-i\vec{\beta}\cdot\vec{X}^{N}}
\end{split}
\label{deltaG}
\end{equation}
for coordinates $\z\left(g\right)$ such that $\z\left(e\right)=0$,
${\cal L}_{i}\z^{j}\left(e\right)=\delta_{i}^{j}$, which has the
property (see \ref{sec:DeltaGroup}) 
\begin{equation}
\delta\left(gh\right)=\mu\left(\z\left(h\right)\right)^{-1}\delta^{d}\left(\z\left(g\right)+\z\left(h\right)\right),\label{deltaGdeltaZ}
\end{equation}
$\mu\left(\z\left(g\right)\right)$ being the Haar measure factor
$dg=\mu\left(\z\left(g\right)\right)d^{d}\z\left(g\right)$. 
\item ${\cal F}$ is an isometry from $L^{2}\left(G\right)$ to $L_{\star}^{2}\left(\g^{*}\right)$
in that it preserves the $L^{2}$ norms (\ref{L2gNorm}) and (\ref{L2Xnorm}):
\begin{equation}
\left\langle \tilde{\psi},\tilde{\psi}'\right\rangle _{\g^{*}}:=\int_{\g^{*}}\frac{d^{d}X}{\left(2\pi\right)^{d}}\left(\overline{\tilde{\psi}\left(X\right)}\star\tilde{\psi}'\left(X\right)\right)=\int_{G}dg\ \overline{\psi\left(g\right)}\psi'\left(g\right)=\left\langle \psi,\psi'\right\rangle _{G},
\end{equation}
so that we may identify $L_{\star}^{2}\left(\g^{*}\right)={\cal F}\left(L^{2}\left(G\right)\right)$. 
\item The $\star$-product is dual to the convolution product $\ast$ on
$G$ under the non-commutative Fourier transform: 
\begin{equation}
\begin{gathered}\tilde{\psi}\star\tilde{\psi}'=\widetilde{\psi\ast\psi'}\\
\text{where}\qquad\psi\ast\psi'\left(g\right)=\int_{G}dh\ \psi\left(gh^{-1}\right)\psi'\left(h\right).
\end{gathered}
\end{equation}
\end{itemize}
Moreover one can see that, in canonical coordinates $k^{i}\left(g\right)\equiv\left(\rho^{i}\left(g\right),\beta^{i}\left(g\right)\right)$,
the Haar measure coincides with the inverse square of the factor (\ref{DufloFactorRho})
coming from the Duflo map: 
\begin{equation}
\mu\left(\vec{\rho}\left(g\right),\vec{\beta}\left(g\right)\right)={\cal A}^{-2}\left(\vec{\rho}\left(g\right),\vec{\beta}\left(g\right)\right).\label{DufloMeasureFeature}
\end{equation}
This remarkable feature reflects the property (\ref{DufloJacobian})
of the Duflo map, and from it some interesting consequences arise
in the characterization of the Fourier transform: 
\begin{itemize}
\item notice first that the property (\ref{DufloMeasureFeature}) implies
that the inner product of two plane waves respect to $\g^{*}$, which
is the group delta function, becomes the pointwise product between
plane waves. Indeed from (\ref{deltaG}), for $g,h\in G$, 
\begin{equation}
\int_{\g^{*}}d^{d}X\ E_{g}\left(X\right)\star E_{h}\left(X\right)=\int_{\g^{*}}d^{d}X\ E_{gh}\left(X\right)=\delta\left(gh\right).
\end{equation}
Using the transformation law (\ref{deltaGdeltaZ}) for canonical coordinates
$z\left(g\right)=\left(\vec{\rho}\left(g\right),\vec{\beta}\left(g\right)\right)$,
\begin{equation}
\begin{split}\delta\left(gh\right)= & \mu^{-1}\left(\z\left(g\right)\right)\delta^{d}\left(\z\left(g\right)+\z\left(h\right)\right)\\
= & \mu^{-1}\left(\z\left(g\right)\right)\int_{\g^{*}}d^{d}X\ e^{i\left(\z\left(g\right)+\z\left(h\right)\right)\cdot X},
\end{split}
\end{equation}
where we used the ordinary representation of the delta function $\delta^{d}\left(\z\left(g\right)+\z\left(h\right)\right)$
with respect to the Lebesgue measure $d^{d}X$. Noticing that the
last expression is non-zero only for $\z\left(h\right)=-\z\left(g\right)$,
and that ${\cal A}\left(-\z\left(g\right)\right)=\overline{{\cal A}\left(\z\left(g\right)\right)}$,
we can rewrite it as 
\begin{equation}
\mu^{-1}\left(\z\left(g\right)\right)\left|{\cal A}^{-2}\left(\z\left(g\right)\right)\right|\int_{\g^{*}}d^{d}X\ {\cal A}\left(\z\left(g\right)\right)e^{i\z\left(g\right)\cdot X}{\cal A}\left(\z\left(h\right)\right)e^{i\z\left(h\right)\cdot X}.\label{IntPlaneWave}
\end{equation}
But from (\ref{DufloMeasureFeature}), for the Duflo map, this implies
\begin{equation}
\delta\left(gh\right)=\int_{\g^{*}}d^{6}X\ E_{g}\left(X\right)\star E_{h}\left(X\right)=\int_{\g^{*}}d^{6}X\ E_{g}\left(X\right)E_{h}\left(X\right).\label{IntDufloPlaneWave}
\end{equation}
This implies that the Duflo $L_{\star}^{2}$ inner product coincides
with the usual $L^{2}$ inner product, and therefore $L_{\star}^{2}\left(\g^{*}\right)\subseteq L^{2}\left(\g^{*}\right)$
as an $L^{2}$ norm-complete vector space for the Duflo map. Indeed,
considering the representation property (\ref{algebraRep-1}) it follows
by linearity from (\ref{IntPlaneWave}) that for a generic quantization
map 
\begin{equation}
\int_{\g^{\star}}d^{d}X\ \tilde{\phi}\left(X\right)\star\tilde{\psi}\left(X\right)=\int_{\g^{\star}}d^{d}X\ \left(\sigma\left(i\vec{\partial}\right)\tilde{\phi}\left(X\right)\right)\tilde{\psi}\left(X\right),
\end{equation}
where $\sigma\left(\z\right)^{-1}=\mu\left(\z\right)\left|{\cal A}\left(\z\right)\right|^{2}$.
Using again (\ref{DufloMeasureFeature}), for the Duflo map last relation
reduces to 
\begin{equation}
\int_{\g^{\star}}d^{d}X\ \tilde{\phi}\left(X\right)\star\tilde{\psi}\left(X\right)=\int_{\g^{\star}}d^{d}X\ \tilde{\phi}\left(X\right)\tilde{\psi}\left(X\right),\label{IntDuflo}
\end{equation}
showing that under integration, the Duflo $\star$-product coincides
with the pointwise product. 
\item A similar feature characterizes the inner product of two plane waves
respect to $G$. It is easy to see from the definition of the Fourier
transform that it corresponds to the non-commutative Dirac delta in
$L_{\star}^{2}\left(\g^{*}\right)$ 
\begin{equation}
\frac{1}{\left(2\pi\right)^{6}}\int_{\text{SL}(2,\C)}dg\ E_{g}\left(X\right)\overline{E_{g}\left(Y\right)}=\delta_{\star}\left(X,Y\right),\label{deltaStar}
\end{equation}
acting as a standard delta distribution with respect to the $\star$-product:
\begin{equation}
\int_{\g^{*}}d^{d}X\ \delta_{\star}\left(X\right)\star\tilde{\psi}\left(X\right)=\int_{\g^{*}}d^{d}X\ \tilde{\psi}\left(X\right)\star\delta_{\star}\left(X\right)=\tilde{\psi}\left(0\right),
\end{equation}
where $\delta_{\star}\left(X\right)=\delta_{\star}\left(X,0\right)$.
But in particular for the Duflo map it follows from (\ref{DufloMeasureFeature})
that 
\begin{equation}
\frac{1}{\left(2\pi\right)^{6}}\int_{\text{SL}(2,\C)}dg\ E_{g}\left(X\right)\overline{E_{g}\left(Y\right)}=\frac{1}{\left(2\pi\right)^{6}}\int_{\text{SL}(2,\C)}d^{3}\vec{\rho}d^{3}\vec{\beta}\ e^{i\vec{\rho}\cdot\left(\vec{X}^{J}-\vec{Y}^{J}\right)+i\vec{\beta}\cdot\left(\vec{X}^{N}-\vec{Y}^{N}\right)},\label{StandardDelta}
\end{equation}
i.e. the inner product of two plane waves reduces to the standard
orthogonality expression in canonical coordinates with respect to
the Lebesgue measure $d^{3}\vec{\rho}d^{3}\vec{\beta}$, and the non-commutative
delta function reduces to the standard delta respect to the pointwise
product, $\delta_{\star}\left(X,Y\right)=\delta\left(X,Y\right)$,
with the only important caveat that the parameters relative to the
compact subgroup have compact range. Specifically, the restriction
on the domain of $\phi$, $\phi\in(-2\pi,2\pi]$ (or $\phi\in(-\pi,\pi]$
for SO(3,1)), implies the domain of $\left(\vec{\rho},\vec{\beta}\right)$
to be restricted by the condition 
\begin{equation}
\sqrt{\frac{1}{2}\left(\left(\rho^{2}-\beta^{2}\right)+\sqrt{\left(\rho^{2}-\beta^{2}\right)^{2}+4\left(\vec{\rho}.\vec{\beta}\right)^{2}}\right)}\in(-2\pi,2\pi]\qquad(\text{or}\ (-\pi,\pi]\ \text{for}\ \text{SO}(3,1)).
\end{equation}
Notice that for the special case in which the boost and rotation parameters
are collinear, i.e. $\vec{\rho}.\vec{\beta}=\rho\beta$, this expression
reduces to a condition on the modulus of the (canonical) rotation
parameter 
\begin{equation}
\rho\in[0,2\pi),\qquad(\text{or}\ [0,\pi)\ \text{for}\ \text{SO}(3,1)).
\end{equation}
Thus, (\ref{StandardDelta}) tells us that the Duflo non-commutative
delta behaves as a standard delta distribution when considering the
commutative space of variables $X$, where however due to the restriction
on the parameter range associated to the compact subgroup, some of
the values of the $X$ spaces are restricted to take discrete values.
This is the expected result, due to the compactness of the corresponding
sections of the conjugate space. In fact, the same feature was pointed
out in previous works concerning non-commutative harmonic analysis
on compact groups~\cite{FreMajDuflo} 
\end{itemize}

\subsection{Relation between the non-commutative Fourier transform and the Fourier
expansion in group unitary irreducible representations}

The non-commutative Fourier transform allows to switch between the
group representation and the algebra representation of the quantum
algebra $\U$. In standard harmonic analysis, however, functions on
the group are expanded in terms of unitary irreducible representations.
In that case a different generalization of the Fourier transform is
considered, consisting in a unitary map from square-integrable functions
$L^{2}\left(G\right)$ to square-integrable functions $L^{2}\left(\hat{G}\right)$
on the Pontryagin dual.

The harmonic analysis on the Lorentz group is developed for instance
in~\cite{ruhl}. It is shown that a function $\psi\left(g\right)\in L^{2}(\text{SL}(2,\C))$
can be expanded in terms of irreducible unitary (infinite dimensional)
representations of the principal series, by means of the Plancherel
decomposition (and its inverse) 
\begin{equation}
\begin{gathered}\begin{gathered}\hat{\psi}_{j_{1}j_{2}q_{1}q_{2}}^{\chi}=\int_{\text{SL}(2,\C)}dg\ D_{j_{1}j_{2}q_{1}q_{2}}^{\chi}(g)\psi\left(g\right),\end{gathered}
\\
\psi\left(g\right)=\tfrac{1}{2}\int_{\R}dr\sum_{m=-\infty}^{+\infty}\left(m^{2}+r^{2}\right)\sum_{j_{1},j_{2}=\left|(1/2)m\right|}^{\infty}\sum_{q_{1}=-j_{1}}^{j_{1}}\sum_{q_{1}=-j_{2}}^{j_{2}}\overline{D_{j_{1}j_{2}q_{1}q_{2}}^{\chi}}\left(g\right)\hat{\psi}_{j_{1}j_{2}q_{1}q_{2}}^{\chi},
\end{gathered}
\label{FourierRuhl}
\end{equation}
where $\chi=\left(m,\r\right)$ labels the representation and, in
the principal series, $r$ is a real continuous parameter, while $m$
takes discrete values. $d\chi=dr\left(m^{2}+\r^{2}\right)$ is the
Plancherel measure, and the Fourier coefficients $\hat{\psi}_{j_{1}j_{2}q_{1}q_{2}}^{\chi}$
are matrix elements of functions $L^{2}\left(\hat{G}\right)$. $D_{j_{1}j_{2}q_{1}q_{2}}^{\chi}\left(g\right)$
are matrix elements of irreducible unitary representations in the
so-called ``canonical basis'', spanned by the set $\phi_{q}^{j}\left(u\right)=\left(2j+1\right)^{1/2}D_{(1/2)m,q}^{j}\left(u\right)$,
$j=\left|\frac{1}{2}m\right|+n$, $n=0,1,2,...$, $-j\leq q\leq j$,
complete in $L_{m}^{2}\left(u\right)$, the space of measurable functions
$\phi\left(u\right)$ on SU(2) covariant on the right cosets of U(1)
as $\phi\left(\gamma u\right)=e^{im\omega}\phi\left(u\right)$, with
$u\in\text{SU\ensuremath{\left(2\right)}}$ and $\gamma=\left(\begin{array}{cc}
\exp\left(i\omega\right) & 0\\
0 & \exp\left(-i\omega\right)
\end{array}\right)$. Here $D_{q_{1}q}^{j_{1}}\left(u_{1}\right)$ define the SU(2) unitary
irreducible representations (Wigner matrices), and the matrices $D_{j_{1}j_{2}q_{1}q_{2}}^{\chi}\left(g\right)$
can be decomposed as 
\begin{equation}
D_{j_{1}j_{2}q_{1}q_{2}}^{\chi}\left(g\right)=\sum_{q}D_{q_{1}q}^{j_{1}}\left(u_{1}\right)D_{qq_{2}}^{j_{2}}\left(u_{2}\right)d_{j_{1}j_{2}q}^{\chi}\left(\eta\right),
\end{equation}
where $u_{1},u_{2}\in\text{SU(2)}$ and $\eta$ is the ``boost''
parameter, and $d_{j_{1}j_{2}q}^{\chi}\left(\eta\right)$ can be written
in the integral representation 
\begin{equation}
\begin{split}d_{j_{1}j_{2}q}^{\chi}\left(\eta\right)= & \left(2j_{1}+1\right)^{1/2}\left(2j_{2}+1\right)^{1/2}\int_{0}^{1}dt\ d_{(1/2)m,q}^{j_{1}}\left(2t-1\right)\\
 & \times d_{(1/2)m,q}^{j_{2}}\left(2t_{d}-1\right)\left[te^{-\eta}+\left(1-t\right)e^{\eta}\right]^{\left(i/2\right)r-1}.
\end{split}
\label{d(eta)}
\end{equation}
Thus, the above makes use of the fact that the space $L_{m}^{2}\left(u\right)$
decomposes into a direct orthogonal sum of $\left(2j+1\right)$-dimensional
Hilbert spaces ${\cal H}_{j}$ carrying each an irreducible representation
of SU(2): $L_{m}^{2}\left(u\right)=\bigoplus_{j=\left|\frac{1}{2}m\right|}^{\infty}{\cal H}_{j}$.

The Fourier transform (\ref{FourierRuhl}) realises the decomposition
of the regular (right and left) representation, carried by the Hilbert
space $L^{2}(\text{SL}(2,\C))$, into irreducible unitary representations
of the principal series. More precisely, consider the kernel $K_{\psi}\left(u_{1},u_{2}|\chi\right)$
of the Fourier transform (\ref{FourierRuhl}) defined by (here $d\mu\left(u\right)$
is the Haar measure for $\text{SU}(2)$) 
\begin{equation}
\hat{\psi}_{j_{1}j_{2}q_{1}q_{2}}^{\chi}=\int d\mu\left(u_{1}\right)d\mu\left(u_{2}\right)\ \overline{\phi_{q_{1}}^{j_{1}}\left(u_{1}\right)}K_{\psi}\left(u_{1},u_{2}|\chi\right)\phi_{q_{2}}^{j_{2}}\left(u_{2}\right),
\end{equation}
and denote, for any representation $\chi$ and $-\chi$ of the principal
series, $L_{m}^{2}(U_{1})_{q}^{j}$ and $L_{m}^{2}(U_{2})_{q}^{j}$
the Hilbert spaces of, respectively, the measurable functions\footnote{The Hilbert spaces $L_{m}^{2}(U_{1})_{q}^{j}$ and $L_{m}^{2}(U_{2})_{q}^{j}$
are equivalent, as it exists an intertwining operator between the
$\chi$ and $-\chi$ representations. } 
\begin{equation}
\begin{gathered}\phi_{\psi}\left(u_{1}|\chi\right)_{q}^{j}=\int d\mu\left(u_{2}\right)\ K_{\psi}\left(u_{1},u_{2}|\chi\right)\phi_{q}^{j}\left(u_{2}\right),\\
\phi_{\psi}\left(u_{2}|\chi\right)_{q}^{j}=\int d\mu\left(u_{1}\right)\ \overline{\phi_{q}^{j}\left(u_{2}\right)}K_{\psi}\left(u_{1},u_{2}|-\chi\right).
\end{gathered}
\end{equation}
One can show that the image of $L^{2}(\text{SL}(2,\C))$ by the Fourier
transform (\ref{FourierRuhl}) can be mapped isometrically into a
Hilbert space ${\cal H}=\int_{\r\geq0}^{\oplus}d\chi\ \bigoplus_{j=\left|\frac{1}{2}m\right|}^{\infty}\bigoplus_{q=-j}^{j}L_{m}^{2}(U_{1})_{q}^{j}$
(or ${\cal H}=\int_{\r\geq0}^{\oplus}d\chi\ \bigoplus_{j=\left|\frac{1}{2}m\right|}^{\infty}\bigoplus_{q=-j}^{j}L_{m}^{2}(U_{2})_{q}^{j}$)
such that a left (or right) translation in $L^{2}(\text{SL}(2,\C))$
generates irreducible unitary representations $\chi$ of the principal
series in $L_{m}^{2}(U_{1})_{q}^{j}$ ($L_{m}^{2}(U_{2})_{q}^{j}$
respectively).

By means of Eqs. (\ref{FourierTransSL2C}) and (\ref{FourierRuhl})
one can obtain the relation between the expansion of a function $L^{2}(\text{SL}(2,\C))$
in terms of non commutative plane waves and the one in terms of irreducible
group representations (Plancherel modes):\textbf{ } 
\begin{equation}
\begin{gathered}\begin{gathered}\hat{\psi}_{j_{1}j_{2}q_{1}q_{2}}^{\chi}=\frac{1}{\left(2\pi\right)^{6}}\int_{\g^{*}}d^{3}X^{J}d^{3}X^{N}\ D_{j_{1}j_{2}q_{1}q_{2}}^{\chi}(X)\star\tilde{\psi}\left(X\right),\\
\tilde{\psi}\left(X\right)=\tfrac{1}{2}\int_{\R}dr\sum_{m=-\infty}^{+\infty}\left(m^{2}+\r^{2}\right)\sum_{j_{1},j_{2}=\left|(1/2)m\right|}^{\infty}\sum_{q_{1}=-j_{1}}^{j_{1}}\sum_{q_{1}=-j_{2}}^{j_{2}}\ \overline{D_{j_{1}j_{2}q_{1}q_{2}}^{\chi}(X)}\hat{\psi}_{j_{1}j_{2}q_{1}q_{2}}^{\chi}
\end{gathered}
\end{gathered}
\end{equation}
\begin{equation}
D_{j_{1}j_{2}q_{1}q_{2}}^{\chi}(X)=\int_{\text{SL}(2,\C)}dgD_{j_{1}j_{2}q_{1}q_{2}}^{\chi}(g)\overline{E_{g}\left(X\right)}.
\end{equation}

\subsection{Comparison with existing results on non-commutative Fourier transform
for non-compact groups}

As we discussed in the introduction, the non-commutative Fourier transform
has played an important role in the context of quantum gravity, and
more specifically in effective models based on the idea of spacetime
non-commutativity. In particular, when the non-commutativity of spacetime
coordinate operators is of Lie algebra type, the associated momentum
space can be described as a curved manifold with a Lie group structure.
The idea of a curved momentum space dates back to M. Born~\cite{BornReciprocity},
as a way to make more symmetric the role of configuration space (generically
curved, in a GR setting) and momentum space, as a possible key to
quantum gravity. The same idea was later formalized more rigorously
in the language of Hopf-algebras (quantum groups)~\cite{MajidFoundation},
through the introduction of a suitable notion of integration. It has
been then further investigated and developed for several examples
of non-commutative spaces (and spacetimes).

Examples of non-commutativity with a non-compact group manifold for
the momenta have been also considered. Particularly relevant for our
analysis is a work investigating a Snyder~\cite{snyder} kind of
non-commutativity in three dimensions~\cite{GirelliLivineSnyder}.
From a group theoretical point of view, it is possible indeed to consider
Snyder non-commutative spacetime to be described by the the quotient
$\so(4,1)/\so(3,1)$, so that the associated momentum space consists
in the homogeneous space $\text{SO}(4,1)/\text{SO}(3,1)$, i.e. de
Sitter space. In~\cite{GirelliLivineSnyder} a ``Euclidean'' version
of three-dimensional Snyder spacetime is considered, where momentum
space consists in the homogeneous space given by the coset $\text{SO}(3,1)/\text{SO}(3)$,
the three-dimensional hyperboloid ${\cal H}_{3}$. The case treated
fits well with the context of the work we have presented here, since
the $\star$-product introduced in Snyder configuration space and
the non-commutative Fourier transform relating it to the curved momentum
space are induced by the group and algebra structure of the Lorentz
group. They have not been directly derived from a choice of quantization
map, but rather postulated at the onset of the analysis. Using our
framework, however, this quantization map can be in principle read
out of the postulated $\star$-product. In particular, the authors
of~\cite{GirelliLivineSnyder} consider two different $\star$-products
(a non-associative and an associative one), corresponding to two different
parametrizations of the $\text{SO}(3)$ and $\text{SO}(3,1)/\text{SO}(3)$
sectors, followed by the application of the symmetric map (\ref{symmMap})
to the two sectors separately, i.e. to a ordering prescription for
the non-commutative operators corresponding to the Cartan decomposition.
It is not clear what are the properties of the chosen quantization
maps nor why they would be preferable to the Duflo map, from a mathematical
perspective. The more detailed relation with our construction will
be studied in future work. We will discuss in the next section the
implementation of our non-commutative Fourier transform, based on
the Duflo quantization map, for a Cartan decomposition of $\text{SO}(3,1)$,
and for the homogeneous space ${\cal H}_{3}\sim\text{SO}(3,1)/\text{SO}(3)$.
This may facilitate a future detailed comparison of the construction
presented in this paper and the one of ~\cite{GirelliLivineSnyder}.

Another much studied example of spacetime non-commutativity associated
to a curved non-compact momentum space related to the Lorentz group,
is that of $\kappa$-Minkowski~\cite{Lukierski-kMink,Zakrzewski-kMink,MajRue-kMink}
(with its associated Hopf-algebra of symmetries, $\kappa$-Poincaré).
In this case, the non-commutativity is of Lie algebra type, and the
associated momentum space is the group manifold $AN_{3}$, corresponding
to half of de Sitter space~\cite{Jurek-kdS}. The properties of the
non-commutative Fourier transform for $\kappa$-Minkowski have been
considered in several studies. The Hopf-algebra point of view, where
a ``time ordering'' prescription is used to define the non-commutative
plane wave is taken in~\cite{MajidOeckl-kFuorier,LukKosMas-kfield,GACMajid-kMink},
and the corresponding quantization map discussed in~\cite{GACagostini-kfield2004}.
The point of view of the group structure of the momentum manifold
is instead central in~\cite{JurekFre-kfield1e2}, and in~\cite{GirLivOriti-kfield}
its formulation in terms of group field theory has been analyzed.
In our language we can understand the non-commutative Fourier transform
discussed in these works, which is the one associated to the so-called
``bicrossproduct'' basis of $\kappa$-Poincaré~\cite{MajRue-kMink},
to correspond to a quantization map consisting in a specific time-ordering
prescription for the non-commutative operators (see again ~\cite{GACagostini-kfield2004},
where alternative choices of orderings are also considered). Once
more, while the physical idea behind the time-ordering is transparent,
it is not clear if this leads to any advantage from the mathematical
perspective of the algebra of quantum observables, compared to the
Duflo map.


\subsection{Aside the non-commutative Fourier transform for the homogeneous space
SL(2,$\C$)/SU(2)}

\label{sec:SL2C/SU2}

Using the results obtained for the Lorentz group, we can describe
as well the properties of the non-commutative Fourier transform for
its homogeneous spaces. We discuss here only the homogeneous space
${\cal H}_{3}\cong\text{SL}(2,\C)/\text{SU}(2)$ (or $\text{SO}(3,1)/\text{SO}(3)$).
This is relevant for the physical applications mentioned in the previous
subsections, and one more physical application of our formalism using
the same homogeneous space will be given in the next section. In fact,
it is also an interesting domain for spin foam models and group field
theory constructions for Lorentzian quantum gravity in 4d \cite{SF,BCmodelLorentzian}.

Consider the decomposition of an element of SL(2,$\C$) 
\begin{equation}
g=kh=\exp\left(i\vec{\mathfrak{b}}\cdot\vec{N}\right)\exp\left(i\vec{\mathfrak{r}}\cdot\vec{J}\right)\label{cartan}
\end{equation}
where $h=\exp\left(i\vec{\mathfrak{r}}\cdot\vec{J}\right)\in$SU(2),
corresponding to the Cartan decomposition. Parametrizing a generic
point on ${\cal H}_{3}$ as 
\begin{equation}
q=q_{0}\id-\vec{q}\cdot\vec{\sigma}\equiv\left(q_{0},\vec{q}\right),
\end{equation}
$q$ is defined by the SL(2,$\C$) action on the origin $q_{a}=\left(1,\vec{0}\right)\equiv\id$:
\begin{equation}
q=gq_{a}g^{\dagger}=kq_{a}k^{\dagger}=\cosh\left(\mathfrak{b}\right)\id-\sinh\left(\mathfrak{b}\right)\hat{\mathfrak{b}}\cdot\vec{\sigma}\equiv\left(\cosh\left(\mathfrak{b}\right),\sinh\left(\mathfrak{b}\right)\hat{\mathfrak{b}}\right).\label{HyperPoint}
\end{equation}
Thus, the quotient space $\text{SL}(2,\C)/\text{SU}(2)$ can be identified
with the coset\footnote{The decomposition works similarly for ${\cal H}_{3}\cong\text{SO}(3,1)/\text{SO}(3)$,
the difference being again in the range of the rotation parameter.
In this case (see also Sec.~\ref{sec:SO31}) a point in the space
${\cal H}_{3}$ is defined by the 4-vector $\left(q_{0},\vec{q}\right)$,
while the action on the origin $q_{0}=\left(1,\vec{0}\right)$ is
given by $q=gq_{a}$ for $g\in\text{SO}(3,1)$. We omit in the following
the distinction between the two cases unless needed.} $gH$, $H\equiv\text{SU}(2)$. Accordingly, in the splitting (\ref{cartan})
the boost element $k$ is a representative of the set of equivalence
classes $[k]:=\left\{ kh,\ \forall h\in\text{SU}(2)\right\} $ defining
$gH$.

So, we can identify functions $f_{{\cal H}_{3}}\left(q\right)$ on
${\cal H}_{3}$ with functions $f\left(gH\right)$ on the coset $\text{SL}(2,\C)/\text{SU}(2)$,
constant on the orbits of $\text{SU}(2)$, through the projection
\begin{equation}
f_{{\cal H}_{3}}\left(q\right)\equiv f\left(gH\right)=\int_{\text{SU}(2)}dh\ f\left(gh\right),\qquad q\in{\cal H}_{3},\quad g\in\text{SL}(2,\C),\quad h\in H\equiv\text{SU}(2)\label{projectorSU2}
\end{equation}
where, for integrable functions on ${\cal H}_{3}$, the Haar measure
$dg$ induces a SL(2,$\C$)-invariant Haar measure $dq$ on ${\cal H}_{3}$
such that~\cite{barutraczka}

\begin{equation}
\int_{\text{SL}(2,\C)}dg\ f\left(g\right)=\int_{{\cal H}_{3}}dq\int_{\text{SU}(2)}dh\ f\left(gh\right).\label{IntegrHom}
\end{equation}

The Haar measure for the decomposition (\ref{cartan}) factorizes
(see \ref{sec:HaarCartan}) into the product of the measure on ${\cal H}_{3}$
and the measure on the subgroup SU(2) 
\begin{equation}
\begin{gathered}dg=dqdh,\qquad q\in{\cal H}_{3},\quad h\in\text{SU}(2),\qquad\text{with}\\
dq=\frac{\sinh^{2}\left(\mathfrak{b}\right)}{\mathfrak{b}^{2}}d^{3}\vec{\mathfrak{b}},\qquad dh=4\frac{\sin^{2}\left(\frac{1}{2}\mathfrak{r}\right)}{\mathfrak{r}^{2}}d^{3}\vec{\mathfrak{r}}
\end{gathered}
\label{HaarCartan}
\end{equation}

The plane wave can be written in Cartan coordinates as 
\begin{equation}
E_{kh}\left(X\right)={\cal A}\left(\vec{\rho}\left(\vec{\mathfrak{b}},\vec{\mathfrak{r}}\right),\vec{\beta}\left(\vec{\mathfrak{b}},\vec{\mathfrak{r}}\right)\right)e^{i\vec{\rho}\left(\vec{\mathfrak{b}},\vec{\mathfrak{r}}\right)\cdot\vec{X}^{J}+i\vec{\beta}\left(\vec{\mathfrak{b}},\vec{\mathfrak{r}}\right)\cdot\vec{X}^{N}},
\end{equation}
where $\left(\vec{\rho}\left(\mathfrak{b},\mathfrak{r}\right),\vec{\beta}\left(\mathfrak{b},\mathfrak{r}\right)\right)$
are given by the BCH formula (\ref{BCHcartan}). Notice now that from
(\ref{BCHcartan}) it follows that 
\begin{equation}
\begin{gathered}\text{for}\quad\mathfrak{b}=0,\qquad\vec{\rho}=\vec{\mathfrak{r}},\quad\phi=\mathfrak{r},\quad\beta=\eta=0,\\
\text{for}\quad\mathfrak{r}=0,\qquad\vec{\beta}=\vec{\mathfrak{b}},\quad\eta=\mathfrak{b},\quad\phi=\rho=0,
\end{gathered}
\end{equation}
so that, from the expression of the Duflo factor (\ref{DufloFactorRho}),
\begin{equation}
\begin{gathered}{\cal A}\left(\vec{\rho}\left(\mathfrak{b}=0,\vec{\mathfrak{r}}\right),\vec{\beta}\left(\mathfrak{b}=0,\vec{\mathfrak{r}}\right)\right)={\cal A}\left(\vec{\mathfrak{r}},0\right)=\frac{\mathfrak{r}^{2}}{4\sin^{2}\left(\tfrac{1}{2}\mathfrak{r}\right)},\\
{\cal A}\left(\vec{\rho}\left(\vec{\mathfrak{b}},\mathfrak{r}=0\right),\vec{\beta}\left(\vec{\mathfrak{b}},\mathfrak{r}=0\right)\right)={\cal A}\left(0,\vec{\mathfrak{b}}\right)=\frac{\mathfrak{b}^{2}}{4\sinh^{2}\left(\tfrac{1}{2}\mathfrak{b}\right)}.
\end{gathered}
\label{DufloFactorCartan}
\end{equation}
Considering that if $h=\id$ $\vec{\mathfrak{r}}=0$ and if $k=\id$
$\vec{\mathfrak{b}}=0$, we can write the plane wave for the particular
group elements $kh|_{h=\id}$ and $kh|_{k=\id}$ as 
\begin{equation}
\begin{gathered}E_{k}\left(X\right)=\frac{\mathfrak{b}^{2}}{4\sinh^{2}\left(\tfrac{1}{2}\mathfrak{b}\right)}e^{i\vec{\mathfrak{b}}\cdot\vec{X}^{N}},\\
E_{h}\left(X\right)=\frac{\mathfrak{r}^{2}}{4\sin^{2}\left(\tfrac{1}{2}\mathfrak{r}\right)}e^{i\vec{\mathfrak{r}}\cdot\vec{X}^{J}},
\end{gathered}
\label{PlaneWaveCartan}
\end{equation}
and, from the property (\ref{EgstarEh}) of $\star$-product, 
\begin{equation}
E_{kh}\left(X\right)=E_{k}\left(X\right)\star E_{h}\left(X\right).
\end{equation}

We can now use Eq. (\ref{projectorSU2}) and the expression for the
(inverse) Fourier transform (\ref{FourierTransSL2C}) to write 
\begin{equation}
f_{{\cal H}_{3}}\left(q\right)\equiv\int_{\text{SU}(2)}dh\ f\left(gh\right)=\frac{1}{\left(2\pi\right)^{6}}\int_{\text{SU}(2)}dh\int_{\g^{*}}d^{d}X\ \overline{E_{gh}\left(X\right)}\star\tilde{f}\left(X\right),
\end{equation}
which can be rewritten, using (notice that $\overline{E_{gh}\left(X\right)}=E_{h^{-1}g^{-1}}\left(X\right)=\overline{E_{h}\left(X\right)}\star\overline{E_{g}\left(X\right)})$,
as 
\begin{equation}
f_{{\cal H}_{3}}\left(q\right)\equiv\frac{1}{\left(2\pi\right)^{6}}\int_{\g^{*}}d^{d}X\int_{\text{SU}(2)}dh\ \overline{E_{h}\left(X\right)}\star\overline{E_{g}\left(X\right)}\star\tilde{f}\left(X\right).\label{fhypStar}
\end{equation}
Using (\ref{PlaneWaveCartan}) and (\ref{HaarCartan}) we notice that
the first term reduces to the ordinary delta function (with compact
range) on the ``rotation'' sector of the algebra 
\begin{equation}
\frac{1}{\left(2\pi\right)^{3}}\int_{\text{SU}(2)}dh\ \overline{E_{h}\left(X\right)}=\frac{1}{\left(2\pi\right)^{3}}\int_{\text{SU}(2)}d^{3}\vec{\mathfrak{r}}\ e^{-i\vec{\mathfrak{r}}\cdot\vec{X}^{J}}=\delta^{3}(\vec{X}^{J})\label{deltaR}
\end{equation}
with $\left|\vec{r}\right|\leq2\pi$ (or $\left|\vec{r}\right|\leq\pi$
for SO(3,1)). We can use now the property of the Duflo $\star$-product
under integral (\ref{IntDuflo}) to eliminate the first $\star$-product
in (\ref{fhypStar}) to obtain 
\begin{equation}
f_{{\cal H}_{3}}\left(q\right)\equiv\frac{1}{\left(2\pi\right)^{3}}\int_{\g^{*}}d^{d}X\ \delta^{3}(\vec{X}^{J})\overline{E_{g}\left(X\right)}\star\tilde{f}\left(X\right).\label{fhypDelta}
\end{equation}
Notice that in this formula, as in the previous ones leading to it,
the $\star$-product used remains the one of $SL(2,\mathbb{C})$,
and all functions are treated as functions on the $T^{*}(SL(2,\mathbb{C}))$
phase space, only appropriately restricted in their dependence on
the domain\footnote{Notice also the slight abuse of notation: the pointwise product is
between the $\star$-function defined by ~\ref{deltaR} (corresponding,
as a function on $\mathbb{R}^{3}$, to the usual delta function),
and the function obtained by $\star$-multiplying the non-commutative
plane wave with the function $\tilde{f}$; this pointwise product
is then evaluated on the point $X$.}.

\section{An example: particle on the hyperboloid}

We will consider now the motion of a free particle on a three-dimensional
(spatial) hyperboloid, as a very simple example of application of
our formalism in a well-understood context. The quantization of the
system can be described in terms of a path-integral formulation. We
will show that, by means of the non-commutative Fourier transform,
the quantum propagator can be easily formulated in the algebra representation,
as its expression assumes, in this simple case, the form one would
expect from the classical action by identifying the (non-commutative)
classical momenta with the algebra elements. In this sense, the main
virtue of the non-commutative Fourier transform and of the algebra
representation is to allow for a description of the quantum system
in which the underlying classical theory is manifest.

The path-integral formulation of quantum mechanics in terms of non-commutative
momenta was described in \cite{OritiRaasakkaSO3} for the case of
SO(3), and the corresponding propagator for a free particle on a sphere
was derived. In dealing with the homogeneous space ${\cal H}_{3}\cong\text{SL}(2,\C)/\text{SU}(2)$
(or SO(3,1)/SO(3)), we will refer to \cite{OritiRaasakkaSO3} for
the details on the general characterization of quantum mechanics in
non-commutative momentum basis.


A remark on notation: we characterize the spatial hyperboloid introducing
a length scale $\l$, corresponding to the radius of curvature, as
$y=\l q=(\l q_{0},\l\vec{q})$, so that the (unit) hyperboloid defined
by (\ref{HyperPoint}) parametrizes the surface $y_{\mu}y^{\mu}=y_{0}^{2}-\vec{y}^{2}=\l^{2}$.
Similarly we define the dimensional coordinates on ${\cal H}_{3}$
$\vec{x}=\l\vec{\mathfrak{b}}$.

\subsection{The propagator in the group representation}

Considering the characterization of the homogeneous space ${\cal H}_{3}\cong\text{SL}(2,\C)/\text{SU}(2)$
of Sec. \ref{sec:SL2C/SU2}, the finite time propagator is defined
as the expectation value in the Hilbert space $L^{2}\left({\cal H}_{3}\right)$
\begin{equation}
K\left(q'',q';t''-t'\right)=\bra{q''}U\left(t''-t'\right)\ket{q'}.\label{propagator}
\end{equation}
where $\ket{q}$ are vectors in $L^{2}\left({\cal H}_{3}\right)$,
and the time evolution operator is defined in terms of the particle
momentum $\boldsymbol{p}$ as 
\begin{equation}
U\left(t\right)=\exp\left(-i\frac{\boldsymbol{p}^{2}t}{2m\hbar}\right).\label{timeEvolution}
\end{equation}
The propagator can be described in the group representation by noticing~\cite{JunkerBohm,JunkerInomata}
that the finite time propagator (\ref{propagator}) can be expressed
as the convolution in the group manifold 
\begin{equation}
K\left(q'',q';t\right)=\lim_{N\rightarrow\infty}\int dg_{1}\cdots\int dg_{N-1}\ K_{\epsilon}\left(g_{0}^{-1}g_{1}\right)\cdots K_{\epsilon}\left(g_{N-1}^{-1}g_{N}\right),\qquad(q_{j}=g_{j}q_{a}g_{j}^{\dagger},\quad q_{N}=q'',\ q_{0}=q')
\end{equation}
of a spherical function $K_{\epsilon}\left(g\right)$ (the short time
propagator), satisfying 
\begin{equation}
K_{\epsilon}\left(g\right)=K_{\epsilon}\left(h_{1}^{-1}gh_{2}\right),\qquad\forall h_{1},h_{2}\in SU(2).\label{symmetry-prop}
\end{equation}

This is our object of interest. We will give its explicit expression
in the following, as well as its decomposition in irreducible representation
of the group, the analogue of the Peter-Weyl decomposition for compact
groups. This will have to be compared to the one obtained via non-commutative
Fourier transform. In order to do so, we have to introduce some more
harmonic analysis on the hyperboloid.

Functions on the group, satisfying the relation ~\ref{symmetry-prop},
i.e. constant in the two-sided coset $HgH$ (for $H=$SU(2)), can
be expanded in terms of zonal spherical functions ${\cal D}_{00}^{l}\left(g\right)$~\cite{JunkerBohm,JunkerInomata,Vilenkin}.
Spherical representations (or class 1 representations) ${\cal D}^{l}$
of $G$ are unitary irreducible representations which have non-null
vectors $\ket{\psi_{\alpha}}$ invariant under transformations of
a subgroup of $G$. SU(2) is a massive subgroup of SL(2,$\C$), meaning
that there is only one vector $\ket{\psi_{0}}$ in any ${\cal D}^{l}$
invariant under SU(2). Then one can define spherical functions 
\begin{equation}
\psi^{l}\left(g\right)=\braket{\psi|\D^{l}\left(g\right)|\psi_{0}}
\end{equation}
such that $\psi^{l}\left(gh\right)=\psi^{l}\left(g\right)$, i.e.
constant in the left coset $gH$, so that $\psi^{l}\left(g\right)$
can be considered as functions on the homogeneous space ${\cal H}_{3}$.
Choosing a basis $\left\{ \ket{u_{i}}\right\} $ such that $\ket{u_{0}}=\ket{\psi_{0}}$,
the matrix elements of ${\cal D}^{l}\left(g\right)$ given by 
\begin{equation}
\D_{m0}^{l}\left(g\right)=\braket{u_{m}|\D^{l}\left(g\right)|u_{0}}
\end{equation}
are called associated spherical functions. A zonal spherical function
$\D_{00}^{l}\left(g\right)$ is thus an associated spherical function
constant on the two-sided coset $HgH$, i.e. such that for any $h,h'\in SU(2)$
$\D_{00}^{l}\left(hgh'\right)=\D_{00}^{l}\left(g\right)$.

For the spherical principal series the parameter $l$ labeling the
representation is continuous and takes the value $l=-1+i\r$ ($\r\geq0$).
In terms of zonal spherical functions the Fourier expansion of a function
$f\left(g\right)$ invariant on the two-sided coset $HgH$ takes the
simpler expression 
\begin{equation}
f\left(g\right)=\frac{1}{\left(2\pi\right)^{2}}\int_{0}^{\infty}d\r\ d_{l}\hat{f}^{l}\D_{00}^{l}\left(g\right),\hspace{1cm}\hat{f}^{l}=\int dg\ f\left(g\right)\D_{00}^{l}\left(g^{-1}\right),
\end{equation}
where $\overline{\D_{00}^{l}}\left(g\right)=\D_{00}^{l}\left(g^{-1}\right)$,
and $d_{l}$ is a ``dimension'' factor defined by the relation 
\begin{equation}
\int_{\text{SL}(2,\C)}\D_{00}^{l}\left(g\right)\D_{00}^{l'}\left(g^{-1}\right)=\frac{1}{d_{l}}\delta\left(l-l'\right).
\end{equation}

The spherical functions are eigenfunctions of the Laplace-Beltrami
operator on ${\cal H}_{3}$, that can be identified with the operator
$\boldsymbol{p}^{2}$ in the time evolution (\ref{timeEvolution}),
and corresponds to a Casimir of the Lorentz algebra, restricted to
its boost components.

The zonal spherical function coincides with the matrix element $d_{000}^{\left(0,2\r\right)}\left(\eta\right)$
of the SL(2,$\C$) representations defined in (\ref{d(eta)}) and
takes the explicit form 
\begin{equation}
\mathcal{D}_{00}^{l}\left(\mathfrak{b}\left(g\right)\right)=d_{000}^{\left(0,2\r\right)}\left(\eta\left(g\right)\right)=\frac{\sin\left(\r\eta\right)}{\r\sinh\left(\eta\right)}.
\end{equation}
The dimension factor can be evaluated to $d_{l}=\r^{2}.$

\
 In the group representation (i.e. configuration space), and using
polar coordinates, the short time propagator takes the form~\cite{JunkerBohm,JunkerInomata}
\begin{equation}
K\left(q,q';\epsilon\right)=\braket{q''|U\left(\epsilon\right)|q'}=\left(\frac{m\l}{2\pi i\hbar\epsilon}\right)^{\frac{3}{2}}\exp\left(\frac{i}{\hbar}\left(\frac{m\l^{2}}{\epsilon}\left(\cosh\Theta-1\right)+\frac{\hbar^{2}\epsilon}{8m\l^{2}}\right)\right),
\end{equation}
where $\Theta$ is the angle of the hyperbolic rotation transforming
$q'$ into $q$: $q=gq'$. As shown in App. \ref{sec:HyperAngle}
the angle $\Theta$ coincides with the modulus of the hyperbolic rotation
$\mathfrak{b}$ in (\ref{HyperPoint}) (and then with $\eta$, when
restricting to the hyperboloid).

\
 The expansion of the propagator in terms of zonal spherical representations
(the analogue of the Peter-Weyl decomposition for compact Lie groups)
takes instead the expression~\cite{JunkerBohm,JunkerInomata} 
\begin{equation}
K\left(q'',q';t\right)=\frac{1}{2\pi^{2}}\int_{0}^{\infty}d\r\exp\left(-i\frac{\hbar\r^{2}}{2m\l^{2}}t\right)\r\frac{\sin\left(x\left(g_{0}^{-1}g_{N}\right)\r/\l\right)}{\sinh\left(x\left(g_{0}^{-1}g_{N}\right)/\l\right)},\label{PropaGroup}
\end{equation}
with $x=\left|\vec{x}\right|$.

This is the expression we would have to compare with the one obtained
via non-commutative Fourier transform.

\subsection{The propagator in the non-commutative momentum representation}

We identify the particle momenta in the algebra representation with
the operators $\hat{X}^{N}\in{\cal Q}\left({\cal A}_{\g^{*}}\right)$
relative to the ``boost'' sector, through the rescaling 
\begin{equation}
\hat{P}=\frac{\hbar}{\l}\hat{X}^{N},
\end{equation}
while we denote $\hat{R}=\frac{\hbar}{l_{0}}\hat{X}^{J}$ the ``momenta''
relative to the rotation sector (which are trivial when we restrict
to the hyperboloid).

Following the construction of~\cite{OritiRaasakkaSO3}, we define
a set of states $\left\{ \ket{\vec{P},\vec{R}}|\vec{P},\vec{R}\in\R_{\star}^{6}\right\} $
in the non-commutative momentum basis, by their inner product with
the group basis 
\begin{equation}
\braket{g|\vec{P},\vec{R}}\equiv E_{g}\left(\vec{P},\vec{R}\right)=\exp_{\star}\left(i\frac{\l}{\hbar}\left(\vec{\beta}\cdot\vec{P}+\vec{\rho}\cdot\vec{R}\right)\right).
\end{equation}
By virtue of properties (\ref{deltaG}) and (\ref{deltaStar}) they
form a complete and orthonormal basis respect to the $\star$-product:
\begin{equation}
\braket{\vec{P},\vec{R}|\vec{P}',\vec{R}'}=\left(\frac{2\pi\hbar}{\l}\right)^{6}\delta_{\star}\left(\vec{P}-\vec{P}',\vec{R}-\vec{R}'\right),\hspace{1cm}\int_{\R_{\star}^{6}}\frac{d^{3}\vec{P}d^{3}\vec{R}}{\left(2\pi\hbar/\l\right)^{6}}\ket{\vec{P},\vec{R}}\star\bra{\vec{P},\vec{R}}=\hat{\id}.\label{momentum-basis}
\end{equation}
On this Hilbert space, corresponding to the algebra representation
of the system, we can define the time evolution operator on the Lorentz
group (\ref{timeEvolution}) to be 
\begin{equation}
U\left(t\right)=-\frac{it}{2m\hbar}\left(\hat{P}^{2}-\hat{R}^{2}\right),
\end{equation}
where $\hat{{\cal C}}=\hat{P}^{2}-\hat{R}^{2}$ is the quadratic Casimir
of the quantum Lorentz algebra. The full propagator is 
\begin{equation}
\begin{split}K_{G}\left(g'',g';t\right)= & \braket{g''|U\left(t\right)|g'}=\\
= & \int_{\R_{\star}^{6}}\frac{d^{3}\vec{P}d^{3}\vec{R}}{\left(2\pi\hbar/\l\right)^{6}}\braket{g''|P,R}\star\braket{P,R|U\left(t\right)|g'}\\
= & \int_{\R_{\star}^{6}}\frac{d^{3}\vec{P}d^{3}\vec{R}}{\left(2\pi\hbar/\l\right)^{6}}E_{g''}\left(\vec{P},\vec{R}\right)\star e_{\star}^{-\frac{i}{2m\hbar}t\sum_{i}\left(P_{i}\star P_{i}-R_{i}\star R_{i}\right)}\star\overline{E_{g'}\left(\vec{P},\vec{R}\right)}
\end{split}
\end{equation}
The expression $\sum_{i}\left(P_{i}\star P_{i}-R_{i}\star R_{i}\right)$
is the quadratic Casimir in the algebra representation ${\cal C}_{\star}$,
obtained with the inverse Duflo map, and reduces to (see (\ref{DufloCasimir}))
\begin{equation}
{\cal C}_{\star}={\cal D}^{-1}\left(\hat{{\cal C}}\right)=\sum_{i}\left(P_{i}\star P_{i}-R_{i}\star R_{i}\right)=\vec{P}^{2}-\vec{R}^{2}+\frac{\hbar^{2}}{\l^{2}}.\label{Cstar}
\end{equation}
Notice that ${\cal C}_{\star}$ $\star$-commutes with functions of
$X$ and we can rewrite the propagator as 
\begin{equation}
K_{G}\left(g'',g';t\right)=\int_{\R_{\star}^{6}}\frac{d^{3}\vec{P}d^{3}\vec{R}}{\left(2\pi\hbar/\l\right)^{6}}e_{\star}^{-\frac{i}{2m\hbar}t{\cal C}_{\star}}\star E_{g''{g'}^{-1}}\left(P,R\right)=K_{G}\left(g=g''{g'}^{-1};t\right)\label{propagatorTotal}
\end{equation}
where we used that $\overline{E_{g}\left(X\right)}=E_{g^{-1}}\left(X\right)$.

Now we exploit relation (\ref{IntegrHom}) to project the propagator
on the homogeneous space ${\cal H}_{3}$ (see also App. \ref{sec:ProjectionH3}
and \cite{DowkerPathHomog}): 
\begin{equation}
K\left(q'',q';t\right)=\int_{SU(2)}dh\ K_{G}\left(g=g''{g'}^{-1};t\right)\label{HyperProject}
\end{equation}
where $q''=g''q_{a}{g''}^{\dagger}$ and $q'=g'q_{a}{g'}^{\dagger}$.
Following the discussion of Sec. \ref{sec:SL2C/SU2}, we rewrite the
plane wave exploiting the splitting (\ref{cartan}) and notice that
the ``boost'' part of the plane wave now takes the form (\ref{DufloFactorCartan}),
in physical coordinates $\vec{x}\left(g\right)=\l\vec{\mathfrak{b}}\left(g\right)$,
\begin{equation}
E_{k}\left(\vec{P},\vec{R}\right)=\frac{\left(x/2\l\right)^{2}}{\sinh^{2}\left(x/2\l\right)}e^{i\vec{x}\cdot\vec{P}/\hbar}.\label{PlaneWaveBoost}
\end{equation}

\noindent Plugging these relations into the propagator, we obtain
\begin{equation}
K\left(q'',q';t\right)=\int_{\R_{\star}^{6}}\frac{d^{3}\vec{P}d^{3}\vec{R}}{\left(2\pi\hbar/\l\right)^{6}}\ e_{\star}^{-\frac{i}{2m\hbar}t\sum_{i}{\cal C}_{\star}}\star E_{k}\left(\vec{P},\vec{R}\right)\star\int_{SU(2)}dh\ E_{h}\left(\vec{P},\vec{R}\right),
\end{equation}
where $\left(\vec{x},\vec{\mathfrak{r}}\right)=\left(\vec{x}\left(g''{g'}^{-1}\right),\vec{\mathfrak{r}}\left(g''{g'}^{-1}\right)\right)$.
Again, the last term is just the delta function $\delta^{3}(\vec{R})$
on the rotation part (\ref{deltaR}), and we get, using the property
(\ref{IntDuflo}) to eliminate the last $\star$-product, 
\begin{equation}
K\left(q'',q';t\right)=\int_{\R_{\star}^{6}}\frac{d^{3}\vec{P}d^{3}\vec{R}}{\left(2\pi\hbar/\l\right)^{6}}\ \delta^{3}(\vec{R})\left(e_{\star}^{-\frac{i}{2m\hbar}t{\cal C}_{\star}}\star E_{k}\left(\vec{P},\vec{R}\right)\right).\label{PropagatorFinal}
\end{equation}
Notice further that due to the $\star$commutavity of the Casimir,
the propagator can be rewritten in terms of a single $\star$-exponential
as
\begin{equation}
K\left(q'',q';t\right)=\int_{\R_{\star}^{6}}\frac{d^{3}\vec{P}d^{3}\vec{R}}{\left(2\pi\hbar/\l\right)^{6}}\ \delta^{3}(\vec{R})\left(e_{\star}^{-\frac{i}{2m\hbar}t\sum_{i}\left(P_{i}\star P_{i}-R_{i}\star R_{i}\right)+\frac{i}{\hbar}\vec{x}\cdot\vec{P}}\right).\label{PropagatorFinalSingle}
\end{equation}
Expression (\ref{PropagatorFinal}) can be simplified noticing that
from the propreties of the Duflo map to preserve the algebra of invariant
polynomials on $\g$, the star-exponential involving the Casimir reduces
to the standard exponential (see (\ref{CasimirStar})), and using
also (\ref{PlaneWaveBoost}) we can rewrite the propagator as 
\begin{equation}
K\left(q'',q';t\right)=\int_{\R_{\star}^{6}}\frac{d^{3}\vec{P}d^{3}\vec{R}}{\left(2\pi\hbar/\l\right)^{6}}\ \delta^{3}(\vec{R})\left(e^{-\frac{i}{2m\hbar}t\left(\vec{P}^{2}-\vec{R}^{2}+\frac{\hbar^{2}}{\l^{2}}\right)}\star e^{i\vec{x}\cdot\vec{P}/\hbar}\right)\frac{\left(x/2\l\right)^{2}}{\sinh^{2}\left(x/2\l\right)}.\label{PropagatorFinal2}
\end{equation}

The expression for the propagator we have obtained, in its form (\ref{PropagatorFinalSingle})
or (\ref{PropagatorFinal2}), is the direct analogue of the commutative
one, and presents the non-commutative algebra variables in the precise
role we expect for the classical momenta in a path integral expression
(here within a Fourier transform to configuration basis). This is
the main result we wanted to show for this simple application of our
formalsim.

As a further remark, we point out that the propagator could be expressed
in terms of only commutative momenta by calculating explictly the
$\star$-product in (\ref{PropagatorFinal2}). We leave the calculation
to a future work. We notice however for the intersted reader that
the calculation will necessarily result in some quantum correction
to the classical expression. Considering that the action of the Casimir
on a plane wave can be rephrased as the Laplacian in terms of Lie
derivatives (see relations (\ref{algebraRep})), we expect that that
the quantum corrections arising from the $\star$-product can be encoded
in an extra muliplicative term ${\cal K}\left(t\hbar/(m\l^{2}),x^{2}/\l^{2}\right)$
consisting is some combination of the dimensionless terms $t\hbar/(m\l^{2})$
and $x^{2}/\l^{2}$, reducing to identity in the classical limit (which
we take to be $\hbar\rightarrow0$ followed by $\l\rightarrow\infty$).
In this case we can apply the delta function $\delta^{3}\left(R\right)$,
and we are left with 
\begin{equation}
K\left(q'',q';t\right)\simeq\int_{\R_{\star}^{3}}\frac{d^{3}\vec{P}}{\left(2\pi\hbar/\l\right)^{3}}\ e^{-\frac{i}{2m\hbar}t\left(\vec{P}^{2}+\frac{\hbar^{2}}{\l^{2}}\right)}e^{i\vec{x}\cdot\vec{P}/\hbar}{\cal K}\left(\frac{t\hbar}{m\l^{2}},\frac{x^{2}}{\l^{2}}\right)\frac{\left(x/2\l\right)^{2}}{\sinh^{2}\left(x/2\l\right)}.\label{PropagatorSolved}
\end{equation}
We can separate the radial and angular part of momenta $\vec{P}=P\hat{n}$
($P=|\vec{P}|$) and, considering that the Casimir does not depend
on the direction of $\vec{P}$, integrate the plane wave on the unit
sphere to get 
\begin{equation}
K\left(q'',q';t\right)\simeq\frac{1}{2\pi^{2}}\int_{0}^{\infty}\frac{\l}{\hbar}dP\ e^{-\frac{i}{2m\hbar}t\left(P^{2}+\frac{\hbar^{2}}{\l^{2}}\right)}{\cal K}\left(\frac{t\hbar}{m\l^{2}},\frac{x^{2}}{\l^{2}}\right)\frac{\l}{\hbar}P\frac{x(g''{g'}^{-1})\sin\left(x(g''{g'}^{-1})P/\hbar\right)}{2\l\sinh^{2}\left(x(g''{g'}^{-1})/\l\right)},\label{propagatorSphere}
\end{equation}
where we also wrote explicitly the dependence on the group element.

Comparing this result with the one for the propagator in the representation
basis (\ref{PropaGroup}), we notice the similarity of the results,
provided one uses the identification $\r=\tfrac{\l}{\hbar}P$. While
the \lq classical\rq appearance of the propagator in the algebra
representation would be maintained for a more complicated quantum
system, eg a particle in an external potential and, more generally,
with a more complicated action, in particular one depending on more
than the quadratic invariant of the momenta, the similarity of the
resulting expression with the expression in representation basis would
be lost. This is to be expected, since harmonic functions are only
eigenstates of the quadratic Casimir, while the states ~\ref{momentum-basis}
are \textit{generalised eigenstates} of all momentum operators. Finally,
we can ascribe the correction terms in the exponential to the choice
of (Duflo) quantization map (see the discussion of a similar term
in the compact case in \cite{OritiRaasakkaSO3}).

To conclude our remark, we consider again expression (\ref{PropagatorSolved})and
take the infinitesimal propagator, for $t\rightarrow\epsilon$, $K\left(q'',q';\epsilon\right)$.
Expanding the factor ${\cal K}\left(\frac{\epsilon\hbar}{m\l^{2}},\frac{x^{2}}{\l^{2}}\right)$
in powers of $\epsilon$ and $x^{2}/\l^{2}$, we rewrite it as 
\begin{equation}
{\cal K}\left(\frac{\epsilon\hbar}{m\l^{2}},\frac{x^{2}}{\l^{2}}\right)\simeq1+{\cal K}_{0}\frac{x^{2}}{\l^{2}}+O\left(\epsilon\right)+O\left(\frac{x^{4}}{\l^{4}}\right),
\end{equation}
where ${\cal K}_{0}$ is some constant factor. Notice that, as expected
in the algebra representation (\ref{algebraRep-1}), we can rewrite
functions of $\vec{x}/\ell$ as differential operators acting on the
plane wave through the substitution $x^{i}\rightarrow-i\hbar\partial/\partial P_{i}$.
We can thus integrate by part the functions involving those factors
and, noticing also that the prefactor ${\cal A}\left(0,\vec{x}/\l\right)$
is at least fourth order in $\vec{x}/\l$, we obtain 
\begin{equation}
K\left(q'',q';\epsilon\right)=\int_{\R_{\star}^{3}}\frac{d^{3}\vec{P}}{\left(2\pi\hbar/\l\right)^{3}}\ e^{i\vec{x}\cdot\vec{P}/\hbar}e^{-\frac{i}{2m\hbar}\epsilon\left(P^{2}+\frac{\hbar^{2}}{\l^{2}}\left(1-6{\cal K}_{0}\right)\right)}+O\left(\epsilon^{2}\right)
\end{equation}
Making explicit its dependence on the group elements we can re-express
the infinitesimal propagator as 
\begin{equation}
K\left(q'',q';\epsilon\right)\simeq\int_{\R^{3}}\frac{d^{3}P}{\left(2\pi\hbar/\l\right)^{3}}\exp\left(\frac{i}{\hbar}\epsilon\left(\frac{\vec{x}(g''{g'}^{-1})}{\epsilon}\cdot\vec{P}-\frac{1}{2m}\left(P^{2}+\frac{\hbar^{2}}{\l^{2}}\left(1-6{\cal K}_{0}\right)\right)\right)\right).
\end{equation}
In the continuum limit, for the infinitesimal propagator we can rewrite
$g'=g\left(t\right)$, $g''=g\left(t+\epsilon\right)$, so that ($\vec{x}(gg^{-1})=0$)
\begin{equation}
\frac{\vec{x}({g'}^{-1}g'')}{\epsilon}=\vec{x}\left(g^{-1}\left(t\right)\frac{dg\left(t\right)}{dt}\right),
\end{equation}
which corresponds to the boost part of the canonical coordinates associated
to the time derivative of the Maurer-Cartan form $\lambda=g^{-1}dg$,
inducing the curved metric in the manifold $g_{ab}=\sum_{c}\lambda_{\ a}^{c}\lambda_{\ b}^{c}$.
Denoting these coordinates as $x^{a}\left(g\right)=x^{i}\lambda_{\ i}^{a}\left(g\right)$
we thus finally get the finite time propagator 
\begin{equation}
K\left(q'',q';t\right)=\int{\cal D}q{\cal D}P\ \exp\left(\frac{i}{\hbar}\int_{t}^{t'}ds\left(\dot{x}^{a}\left(g\left(s\right)\right)\vec{P}_{a}-\frac{1}{2m}\left(P^{2}+\frac{\hbar^{2}}{\l^{2}}\left(1-6{\cal K}_{0}\right)\right)\right)\right).
\end{equation}
The propagator takes the explicit form of a path-integral whose action
appearing at the exponential coincides with the classical (Hamiltonian)
action, confirming what we had anticipated looking at the infinitesimal
propagator in the algebra representation. One can notice the quantum
correction in the action (whose explicit form should be calculated
solving the $\star$-product in (\ref{PropagatorFinal2})), as a shift
in the energy, which characterizes the Duflo quantization map we have
chosen. Similar corrections had been found in the Euclidean case,
with a different quantization map, in \cite{OritiRaasakkaSO3}.

\section{Conclusions}

We have defined a non-commutative algebra representation for quantum
systems whose phase space is the cotangent bundle of the Lorentz group
$T*SL(2,\mathbb{C})$, and the non-commutative Fourier transform ensuring
the unitary equivalence with the standard group representation. Our
construction, following the general template presented in \cite{GuOrRaNCFT},
is \textit{from first principles} in the sense that all the structures
are derived from the single initial input of a choice of quantization
map for the classical system. Our specific construction corresponds
to the choice of the Duflo quantization map, a choice motivated by
the special mathematical properties of this map as well as by the
interesting physical applications of the same, as we discussed in
the text.

\
 We have left all possible physical applications (beside the simple
case of a point particle) of our construction aside, in this paper.
However, we believe that our results could be of considerable impact
in this direction. We have in mind in particular the application to
quantum gravity, which can take two parallel paths. The first is in
the context of model building for Lorentzian 4d quantum gravity, within
the group field theory and spin foam formalisms, so far quite limited
(model building based on the Duflo map in the Riemannian context is
discussed in \cite{daniele-marco}). In this context, Lorentzian model
building should proceed alongside a more careful investigation of
causal (or, better, \lq pre-causal\rq) properties of the resulting
models, possibly inspired by the early work \cite{daniele-etera-causalitySF}.
The second is in the context of non-commutative spacetime field theories,
where on the one hand our construction can offer a more mathematically
solid ground for the construction of effective quantum gravity models,
while on the other hand the possible phenomenological implications
of the mathematical peculiarities of the Duflo map could be identified.


\appendix

\section{Some properties of the Lorentz group parametrization}

\subsection{Explicit expression for the SO(3,1) matrices and the relation between
SO(3,1) and SL(2,C) representations.}

\label{sec:SO31explicit}

The explicit expression of $\Lambda\left(\vec{\rho},\vec{\beta}\right)$
can be obtained directly from (\ref{Lambda(a)}) or evaluating the
matrix exponential (\ref{Lambda(J,N)}), and it is given by 
\begin{equation}
\begin{gathered}\Lambda_{\ 0}^{0}=\frac{1}{\phi^{2}+\eta^{2}}\left(\frac{1}{2}\left(\phi^{2}+\eta^{2}+\vec{\rho}^{2}+\vec{\beta}^{2}\right)\cosh\eta+\frac{1}{2}\left(\phi^{2}+\eta^{2}-\vec{\rho}^{2}-\vec{\beta}^{2}\right)\cos\phi\right),\\
\Lambda_{\ i}^{0}=\frac{1}{\phi^{2}+\eta^{2}}\left(-\rho_{i}\left(\phi\sinh\eta-\eta\sin\phi\right)-\beta_{i}\left(\eta\sinh\eta+\phi\sin\phi\right)+\epsilon_{ijk}\rho_{j}\beta_{k}\left(\cosh\eta-\cos\phi\right)\right),\\
\Lambda_{\ 0}^{i}=\frac{1}{\phi^{2}+\eta^{2}}\left(-\rho_{i}\left(\phi\sinh\eta-\eta\sin\phi\right)-\beta_{i}\left(\eta\sinh\eta+\phi\sin\phi\right)-\epsilon_{ijk}\rho_{j}\beta_{k}\left(\cosh\eta-\cos\phi\right)\right)\\
\begin{split}\Lambda_{\ j}^{i}= & \frac{1}{\phi^{2}+\eta^{2}}\Bigg(\delta_{ij}\left(\frac{1}{2}\left(\phi^{2}+\eta^{2}-\rho^{2}-\beta^{2}\right)\cosh\eta+\frac{1}{2}\left(\phi^{2}+\eta^{2}+\rho^{2}+\beta^{2}\right)\cos\phi\right)\\
 & +\left(\beta_{i}\beta_{j}+\rho_{i}\rho_{j}\right)\left(\cosh\eta-\cos\phi\right)+\epsilon_{ijk}\left(\beta_{k}\left(\phi\sinh\eta-\eta\sin\phi\right)-\rho_{k}\left(\eta\sinh\eta+\phi\sin\phi\right)\right)\Bigg),
\end{split}
\end{gathered}
\end{equation}
where the relation between $\phi$, $\eta$, $\vec{\rho}$, $\vec{\beta}$,
are given by (\ref{phietarhobeta}). This shows the relation between
canonical coordinates on SL(2,$\C$) and SO(3,1). As mentioned in
the main text, the parameters domain is different for the two groups:
the multivaluedness of the logarithmic map (the inverse of the exponential
map) is determined by the periodicity of the compact subgroup of rotations.
When $\eta=0$ ($|\vec{\rho}|=\phi$, $\vec{\beta}=0$), the group
SL(2,$\C$) (see (\ref{SL(2,C)complexRotation})) reduces to SU(2)
and the matrix $\Lambda$ represents a pure rotation 
\begin{equation}
\Lambda\left(\vec{\rho},\vec{\beta}=0\right)=\left(\begin{array}{cc}
1 & 0_{3}\\
0_{3} & {\cal R}
\end{array}\right),
\end{equation}
where 
\begin{equation}
{\cal R}_{ij}=\delta_{ij}\cos\rho+\frac{\rho_{i}\rho_{j}}{\rho^{2}}\left(1-\cos\rho\right)-\epsilon_{ijk}\frac{\rho_{k}}{\rho}\sin\rho,
\end{equation}
so that 
\begin{equation}
\Lambda\left(\left|\vec{\rho}\right|=0,\vec{\beta}=0\right)=\Lambda\left(\left|\vec{\rho}\right|=2\pi,\vec{\beta}=0\right)=\id_{4}.
\end{equation}
While for the SU(2) subgroup of SL(2,$\C$) we can see by setting
$\eta=\vec{\beta}=0$ in (\ref{SL(2,C)complexRotation}) that 
\begin{equation}
a\left(\rho=0,\beta=0\right)=a\left(\rho=4\pi,\beta=0\right)=\id_{2},\qquad a\left(\rho=2\pi,\beta=0\right)=-\id_{2}.
\end{equation}
This shows that, analogously to the relation between SO(3) and SU(2),
SL(2,$\C$) ``covers twice'' SO(3,1), manifesting the isomorphism
$\text{SO}(3,1)\simeq\text{SL(2,}\C)/\left\{ \id,-\id\right\} $.

When $\phi=0+2n\pi$ and $\eta\neq0$ ($\rho=0$, $\beta=\eta$),
the matrix $\Lambda$ represents a pure boost 
\begin{equation}
\Lambda\left(\vec{\rho}=0,\vec{\beta}\right)=\left(\begin{array}{cc}
{\cal B}_{00} & {\cal B}_{0i}\\
{\cal B}_{0i} & {\cal B}_{ij}
\end{array}\right)
\end{equation}
where 
\begin{equation}
\begin{gathered}{\cal B}_{00}=\cosh\beta,\qquad{\cal B}_{0i}=-\frac{\beta_{i}}{\beta}\sinh\beta,\\
{\cal B}_{ij}=\delta_{ij}+\frac{\beta_{i}\beta_{j}}{\vec{\beta}^{2}}\left(\cosh\beta-1\right).
\end{gathered}
\end{equation}

\subsection{Branch cuts for the canonical coordinates\label{sec:branchCuts}}

When $\phi\neq0+2n\pi$ or $\eta\neq0$, $a$ can be diagonalized
and $\Lambda$ can be put in a normal form: 
\begin{equation}
a=a\delta a^{-1},\qquad\delta=\left(\begin{array}{cc}
e^{\left(\phi+i\eta\right)/2} & 0\\
0 & e^{-\left(\phi+i\eta\right)/2}
\end{array}\right),
\end{equation}
\begin{equation}
\Lambda\left(a\right)=\Lambda\left(a\right)\Lambda\left(\delta\right)\Lambda\left(a^{-1}\right),\qquad\Lambda\left(\delta\right)=\left(\begin{array}{cccc}
\cosh\eta & 0 & 0 & \sinh\eta\\
0 & \cos\phi & \sin\phi & 0\\
0 & -\sin\phi & \cos\phi & 0\\
\sinh\eta & 0 & 0 & \cosh\eta
\end{array}\right).
\end{equation}
Then $\Lambda$ belongs to an equivalent class of elements $\gamma\left(\phi,\eta\right)$
corresponding to the first of (\ref{conjugClassesSL(2,C)}). If we
restrict $0\leq\phi<2\pi$, each of these classes corresponds to a
unique complex rotation angle, while one class $\gamma\left(\phi,\eta\right)$
of elements of $\Lambda\in\text{SO(3,1)}$ corresponds to two classes
$\gamma\left(\phi,\eta\right)$ and $\gamma\left(\phi+2\pi,\eta\right)$
of elements of SL(2,$\C$).

If $\phi=0+2n\pi$ and $\eta=0$, so that $\vec{\zeta}^{2}=0$, and
if also $\sum_{j}\left|\zeta_{j}\right|^{2}=0$ ($\vec{\rho}=0$,
$\vec{\beta}=0$), then $a$ belongs to $\gamma\left(0,0\right)$.
The second class in (\ref{conjugClassesSL(2,C)}) can be obtained
when $\phi=0$, $\eta=0$, so that $\vec{\zeta}^{2}=0$, but $\sum_{j}\left|\zeta_{j}\right|^{2}>0$
($\vec{\rho}^{2}=\vec{\beta}^{2}$, $\vec{\rho}\cdot\vec{\beta}=0$
but $\vec{\rho}^{2}+\vec{\beta}^{2}>0$)\footnote{For instance the representative of the second class in (\ref{conjugClassesSL(2,C)})
can be obtained by setting $\beta_{1}=-1$, $\rho_{2}=1$ and all
the other components to zero.}. The third class in (\ref{conjugClassesSL(2,C)}) cannot be obtained~\cite{rossmann,hofmanExp}
with the exponential map. Indeed the group SL(2,$\C$) is not exponential.
But one can show that it is weakly exponential. In order to do so
consider the inverse map 
\begin{equation}
\zeta_{j}=\frac{2}{i}\frac{\log(a_{0}+(a_{0}^{2}-1)^{\frac{1}{2}})}{(a_{0}^{2}-1)^{\frac{1}{2}}}a_{j}.\label{inverseMapzeta}
\end{equation}
The study of the branch points of this function shows that, restricting
it to its principal values, the complex rotation vector is holomorphic
in $a_{0}$ except for a branch cut extending on the real axis of
$a_{0}$ from -1 to $-\infty$. Indeed the function in (\ref{inverseMapzeta})
is single valued for all complex $a_{0}$. For $a_{0}$ real with
$-1\leq a_{0}\leq1$ the numerator in (\ref{inverseMapzeta}) is nothing
but $\arccos\left(a_{0}\right)=\frac{1}{i}\log(a_{0}+i\sqrt{1-a_{0}^{2}})$,
which, extended to complex numbers, has branch cuts on the real axis
for $a_{0}<-1$ and $a_{0}>1$. The interval $-1\leq a_{0}\leq1$
(and the imaginary axis of $a_{j}$) is realized by $\eta=0$, $\phi\in\left(-2\pi,2\pi\right]$
, i.e. by the SU(2) subgroup of rotations. The half-line $a_{0}$
real with $a_{0}>1$ is taken by pure boosts $\phi=0,$ $\eta\neq0$,
and we can write the numerator as $\text{arccosh}\left(a_{0}\right)=\log(a_{0}+\sqrt{a_{0}+1}\sqrt{a_{0}-1})$,
which, extended to complex $a_{0}$, has a branch cut on the real
axis for $a_{0}<1$. The branch cut $a_{0}<-1$ is given by the composition
of these two functions. The branch point $a_{0}=-1$ corresponds to
the third class in (\ref{conjugClassesSL(2,C)}). Thus, except for
the branch cut, the canonical coordinates provided by the exponential
map, represented by the complex rotation vector $\vec{\zeta}=\vec{\rho}+i\vec{\beta}$,
are holomorphic functions parametrizing the whole SL(2,$\C$) group.

\subsection{Haar measure for SL(2,$\C$)\label{sec:HaarSL2C}}

Considering an element 
\begin{equation}
a=\left(\begin{array}{cc}
\alpha & \beta\\
\gamma & \delta
\end{array}\right)\in\text{SL}(2,\C),
\end{equation}
the Haar measure is defined~\cite{barutraczka,ruhl} as (the factor
2 is arbitrary) 
\begin{equation}
dg=\frac{2}{\left|\delta\right|^{2}}d\beta d\gamma d\delta d\beta^{*}d\gamma^{*}d\delta^{*},
\end{equation}
and has the invariance property 
\begin{equation}
dg=d\left(hg\right)=d\left(gh\right)=dg^{-1}.
\end{equation}

In the parametrization (\ref{SL(2,C)element}) evaluating the Jacobian
\begin{equation}
{\cal J}\left(\left(\beta,\gamma,\delta,\beta^{*},\gamma^{*},\delta^{*}\right)\rightarrow\left(\vec{a},\vec{a}^{*}\right)\right)=4\left|\frac{a_{3}-a_{0}}{a_{0}}\right|^{2}
\end{equation}
one gets the measure 
\begin{equation}
dg=\frac{8}{\left|a_{0}\left(\vec{a}\right)\right|^{2}}d^{3}\vec{a}d^{3}\vec{a}^{*}\label{Haar-a}
\end{equation}
where 
\begin{equation}
a_{0}\left(\vec{a}\right)=\sqrt{1+\vec{a}^{2}}.
\end{equation}
The Jacobian to complex vector $\left(\zeta,\zeta^{*}\right)$ (\ref{SL(2,C)complexVector})
can be evaluated to 
\begin{equation}
{\cal J}\left(\left(\vec{a},\vec{a}^{*}\right)\rightarrow(\vec{\zeta},\vec{\zeta}^{*})\right)=\left|\frac{\sin\left(\tfrac{1}{2}\zeta\right)\sin\left(\zeta\right)}{4\zeta^{2}}\right|^{2},
\end{equation}
from which 
\begin{equation}
dg=2\frac{\left|\sin^{4}\left(\tfrac{1}{2}\zeta\right)\right|}{\left|\zeta^{4}\right|}d^{3}\zeta d^{3}\zeta^{*}.
\end{equation}
Finally, in terms of the real canonical coordinates $\left(\rho,\beta\right)$
(\ref{complexRotation}) the Haar measure takes the form (${\cal J}\left((\zeta,\zeta^{*})\rightarrow\left(\rho,\beta\right)\right)=8$)
\begin{equation}
dg=\left(4\frac{\cosh^{2}\left(\tfrac{1}{2}\eta\right)\sin^{2}\left(\tfrac{1}{2}\phi\right)+\sinh^{2}\left(\tfrac{1}{2}\eta\right)\cos^{2}\left(\tfrac{1}{2}\phi\right)}{\phi^{2}+\eta^{2}}\right)^{2}d^{3}\vec{\rho}d^{3}\vec{\beta}.\label{HaarSL2C}
\end{equation}
Notice that the Haar measure in terms of canonical coordinates manifests
the property (\ref{DufloJacobian}) of the Duflo map.

\subsection{Haar measure in Cartan decomposition\label{sec:HaarCartan}}

Considering the representation (\ref{complexRotation}) and (\ref{cartan})
we obtain the relations 
\begin{equation}
\begin{gathered}\cos\left(\tfrac{1}{2}\zeta\right)=\cosh\left(\tfrac{1}{2}\left|\vec{\mathfrak{b}}\right|\right)\cos\left(\tfrac{1}{2}\left|\vec{\mathfrak{r}}\right|\right)-i\sinh\left(\tfrac{1}{2}\left|\vec{\mathfrak{b}}\right|\right)\sin\left(\tfrac{1}{2}\left|\vec{\mathfrak{r}}\right|\right)\hat{\mathfrak{b}}\cdot\hat{\mathfrak{r}},\\
\sin\left(\tfrac{1}{2}\zeta\right)\hat{\zeta}=\cosh\left(\tfrac{1}{2}\left|\vec{\mathfrak{b}}\right|\right)\sin\left(\tfrac{1}{2}\left|\vec{\mathfrak{r}}\right|\right)\hat{\mathfrak{r}}+i\sinh\left(\tfrac{1}{2}\left|\vec{\mathfrak{b}}\right|\right)\cos\left(\tfrac{1}{2}\left|\vec{\mathfrak{r}}\right|\right)\hat{\mathfrak{b}}-i\sinh\left(\tfrac{1}{2}\left|\vec{\mathfrak{b}}\right|\right)\sin\left(\tfrac{1}{2}\left|\vec{\mathfrak{r}}\right|\right)\hat{\mathfrak{b}}\wedge\hat{\mathfrak{r}},
\end{gathered}
\label{BCHcartan}
\end{equation}
where $\vec{\zeta}=\vec{\rho}+i\vec{\beta}$ and $\zeta^{2}=\vec{\zeta}^{2}=\left(\phi+i\eta\right)^{2}$.
Eq.~(\ref{BCHcartan}) is nothing but the Baker-Campbell-Hausdorff
(BCH) formula for the Cartan group element~(\ref{cartan}): 
\begin{equation}
g=\exp\left(i\vec{\rho}\cdot\vec{J}+i\vec{\beta}\cdot\vec{N}\right)=\exp\left(i\vec{\mathfrak{b}}\cdot\vec{N}\right)\exp\left(i\vec{\mathfrak{r}}\cdot\vec{J}\right).
\end{equation}

Defining the quantities 
\begin{equation}
\begin{gathered}\vec{Z}=\sin\left(\tfrac{1}{2}\zeta\right)\frac{\vec{\zeta}}{\zeta},\\
\vec{\mathfrak{B}}=\sinh\left(\tfrac{1}{2}\left|\vec{\mathfrak{b}}\right|\right)\hat{\mathfrak{b}},\\
\vec{\mathfrak{R}}=\sin\left(\tfrac{1}{2}\left|\vec{\mathfrak{r}}\right|\right)\hat{\mathfrak{r}},
\end{gathered}
\end{equation}
we get 
\begin{equation}
\sqrt{1-\vec{Z}^{2}}=\sqrt{1+\vec{\mathfrak{B}}^{2}}\sqrt{1-\vec{\mathfrak{R}}^{2}}-i\vec{\mathfrak{B}}\cdot\vec{\mathfrak{R}}
\end{equation}
\begin{equation}
\vec{Z}=\sqrt{1+\vec{\mathfrak{B}}^{2}}\vec{\mathfrak{R}}+i\sqrt{1-\vec{\mathfrak{R}}^{2}}\vec{\mathfrak{B}}+i\vec{\mathfrak{R}}\wedge\vec{\mathfrak{B}}
\end{equation}
In terms of the $\vec{Z}$ coordinates the measure is (see (\ref{Haar-a})
and consider that $\vec{a}=i\vec{Z}$) 
\begin{equation}
dg=\frac{8}{\left|\sqrt{1-\vec{Z}^{2}}\right|^{2}}d^{3}\vec{Z}d^{3}\vec{Z}^{*}
\end{equation}
The Jacobian of transformation from $\vec{Z}$ to $\left(\vec{\mathfrak{B}},\vec{\mathfrak{R}}\right)$
is 
\begin{equation}
\left|\frac{d^{3}\vec{Z}d^{3}\vec{Z}^{*}}{d^{3}\vec{\mathfrak{R}}d^{3}\vec{\mathfrak{B}}}\right|=8\frac{\sqrt{1+\vec{\mathfrak{B}}^{2}}}{\sqrt{1-\vec{\mathfrak{R}}^{2}}}\left(\left(1-\vec{\mathfrak{R}}^{2}\right)\left(1+\vec{\mathfrak{B}}^{2}\right)+\left(\vec{\mathfrak{R}}\cdot\vec{\mathfrak{B}}\right)^{2}\right)
\end{equation}
The measure then becomes 
\begin{equation}
dg=64\frac{\sqrt{1+\vec{\mathfrak{B}}^{2}}}{\sqrt{1-\vec{\mathfrak{R}}^{2}}}d^{3}\vec{\mathfrak{R}}d^{3}\vec{\mathfrak{B}}
\end{equation}
We can now rewrite it in terms of $\vec{\mathfrak{b}}$ and $\vec{\mathfrak{r}}$
coordinates. Considering the Jacobian of transformation (apart from
numerical factors) 
\begin{equation}
\begin{gathered}d^{3}\vec{\mathfrak{R}}=\frac{\sin\left(\tfrac{1}{2}\left|\vec{\mathfrak{r}}\right|\right)\sin\left(\left|\vec{\mathfrak{r}}\right|\right)}{4\left|\vec{\mathfrak{r}}\right|^{2}}d^{3}\vec{\mathfrak{r}},\\
d^{3}\vec{\mathfrak{B}}=\frac{\sinh\left(\tfrac{1}{2}\left|\vec{\mathfrak{b}}\right|\right)\sinh\left(\left|\vec{\mathfrak{b}}\right|\right)}{4\left|\vec{\mathfrak{b}}\right|^{2}}d^{3}\vec{\mathfrak{b}},
\end{gathered}
\end{equation}
the measure becomes 
\begin{equation}
dg=4\frac{\sinh^{2}\left|\vec{\mathfrak{b}}\right|}{\left|\vec{\mathfrak{b}}\right|^{2}}\frac{\sin^{2}\left(\tfrac{1}{2}\left|\vec{\mathfrak{r}}\right|\right)}{\left|\vec{\mathfrak{r}}\right|^{2}}d^{3}\vec{\mathfrak{r}}d^{3}\vec{\mathfrak{b}}\label{MeasureCartan}
\end{equation}

\subsection{Delta on the group in canonical coordinates\label{sec:DeltaGroup}}

Consider that for canonical coordinates in $G$ 
\begin{equation}
z\left(gh\right)={\cal B}\left(z\left(g\right),z\left(h\right)\right),
\end{equation}
where ${\cal B}\left(z\left(g\right),z\left(h\right)\right)$ is the
given by the BCH formula respect to the Lie algebra $\g$. Indeed
\begin{equation}
gh=e^{iz\left(g\right)\cdot x}e^{iz\left(h\right)\cdot x}=e^{i{\cal B}\left(z\left(g\right),z\left(h\right)\right)\cdot x},\qquad g,h\in G,\quad x\in\g.
\end{equation}
Since $z\left(g^{-1}\right)=-z\left(g\right)$, $z\left(e\right)=0$,
\begin{equation}
{\cal B}\left(z\left(g\right),z\left(h\right)\right)=0\qquad\text{for}\qquad z\left(h\right)=-z\left(g\right).
\end{equation}
The delta transforms with the inverse of the Jacobian of transformation
\begin{equation}
\delta\left(gh\right)=\delta^{d}\left({\cal B}\left(z\left(g\right),z\left(h\right)\right)\right)=\left|\frac{d^{d}{\cal B}\left(z\left(g\right),z\left(h\right)\right)}{d^{d}z\left(g\right)}\right|_{z\left(h\right)=-z\left(g\right)}^{-1}\delta^{d}\left(z\left(g\right)+z\left(h\right)\right).
\end{equation}
But from the invariance of the Haar measure $d\left(gh\right)=dg$,
so that 
\begin{equation}
dg=d^{d}z\left(g\right)\mu\left(z\left(g\right)\right)=d^{d}z\left(gh\right)\mu\left(z\left(gh\right)\right)
\end{equation}
it follows that the Jacobian is nothing but 
\begin{equation}
\left|\frac{d^{d}{\cal B}\left(z\left(g\right),z\left(h\right)\right)}{d^{d}z\left(g\right)}\right|=\frac{\mu\left(z\left(g\right)\right)}{\mu\left({\cal B}\left(z\left(g\right),z\left(h\right)\right)\right)},
\end{equation}
and we find, since $\mu\left(z\left(g\right)=0\right)=\mu\left(e\right)=1$,
\begin{equation}
\delta\left(gh\right)=\mu^{-1}\left(z\left(g\right)\right)\delta^{d}\left(z\left(g\right)+z\left(h\right)\right).\label{eq:TransformationMeasure}
\end{equation}

\section{Some calculation for the Duflo quantization map and the associated
$\star$-product}

\subsection{Calculation of the Duflo factor\label{sec:DufloFactor}}

The Duflo function (\ref{j(X)}) can be evaluated explicitly making
use of the identity 
\begin{equation}
\frac{\sinh\left(\frac{1}{2}x\right)}{\frac{1}{2}x}=\exp\left(\sum_{n\geq1}\frac{B_{2n}}{2n\left(2n\right)!}x^{2n}\right),\label{bernoulli}
\end{equation}
with $B_{2n}$ Bernoulli numbers, so that, from the property of the
determinant 
\begin{equation}
\det\left(\exp\left(A\right)\right)=\exp\left(\text{Tr}A\right),
\end{equation}
we get 
\begin{equation}
j\left(x\right)=\exp\left(\sum_{n\geq1}\frac{B_{2n}}{2n\left(2n\right)!}\text{Tr}\left(\text{ad}_{x}\right)^{2n}\right).
\end{equation}

In order to simplify the calculations we adopt the following standard
redefinition of the Lorentz generators: 
\begin{equation}
\vec{L}=\frac{1}{2}\left(\vec{J}-i\vec{N}\right),\qquad\vec{R}=\frac{1}{2}\left(\vec{J}+i\vec{N}\right),\label{eq:ChiralLorentzMap}
\end{equation}
satisfying the brackets 
\begin{equation}
\left[L_{i},L_{j}\right]=i\epsilon_{ijk}L_{k},\qquad\left[R_{i},R_{j}\right]=i\epsilon_{ijk}R_{k},\qquad\left[L_{i},R_{j}\right]=0.\label{eq:algebraLR}
\end{equation}
We have thus split the algebra in two mutually commuting sets of $\su$(2)
generators $L_{i}$ and $R_{i}$, in terms of which a group element
(\ref{gEXP}) takes the form 
\begin{equation}
g=\exp\left(i\vec{\zeta}\cdot\vec{L}\right)\exp\left(i\vec{\zeta}^{*}\cdot\vec{R}\right)
\end{equation}
where $\zeta_{i}=\rho_{i}+i\beta_{i}$ as above. Defining a generic
element of $\g$ in this basis as $x=x^{I}E_{I}=x_{L}^{i}L_{i}+x_{R}^{i}R_{i}$,
for $I=1,\dots,6,$ with $E_{I}=L_{I}$ for $I=1,2,3$, $E_{I}=R_{I-3}$
for $I=4,5,6$, the adjoint representation is given by the matrix
$\left(\ad_{X}\right)_{J}^{K}=x^{I}c_{IJ}^{\ \ K}$, where $c_{IJ}^{\ \ K}$
are the structure constant of $\g$ given by (\ref{eq:algebraLR})
as $c_{IJ}^{\ \ K}=i\epsilon_{IJ}^{\ \ K}$ for $I,J,K=1,2,3$, while
$c_{IJ}^{\ \ K}=i\epsilon_{I-3J-3}^{\ \ K-3}$ for $I,J,K=4,5,6$.
It follows by direct computation that 
\begin{equation}
\ad_{X}\ad_{Y}=\left\{ \left(\text{ad}_{x}\right)_{J}^{I}\left(\text{ad}_{y}\right)_{I}^{K}\right\} =\left(\begin{array}{cc}
\left\{ \delta_{j}^{k}\vec{x}_{L}\cdot\vec{y}_{L}-x_{Lj}y_{L}^{k}\right\}  & \mathbf{0}_{3\times3}\\
\mathbf{0}_{3\times3} & \left\{ \delta_{j}^{k}\vec{x}_{R}\cdot\vec{y}_{R}-x_{Rj}y_{R}^{k}\right\} 
\end{array}\right),
\end{equation}
so that, with the notation $x=\sqrt{\vec{x}\cdot\vec{x}}$ ($=\left|\vec{x}\right|$
if $\vec{x}$ is real) 
\begin{equation}
\Tr\left(\ad_{x}\ad_{x}\right)=2x^{2}=2x_{L}^{2}+2x_{R}^{2}
\end{equation}
With a similar calculation one finds that 
\begin{equation}
\Tr\left(\ad_{x}^{2n}\right)=2x_{L}^{2n}+2x_{R}^{2n},
\end{equation}
and finally 
\begin{equation}
\begin{split}j\left(x\right)= & \exp\left(2\sum_{n\geq1}\frac{B_{2n}}{2n\left(2n\right)!}\left(x_{L}^{2n}+x_{\text{R}}^{2n}\right)\right)\\
= & \exp\left(2\sum_{n\geq1}\frac{B_{2n}}{2n\left(2n\right)!}x_{L}^{2n}\right)\exp\left(2\sum_{n\geq1}\frac{B_{2n}}{2n\left(2n\right)!}x_{R}^{2n}\right)\\
= & 16\frac{\sinh^{2}\left(\tfrac{1}{2}x_{L}\right)}{x_{L}^{2}}\frac{\sinh^{2}\left(\tfrac{1}{2}x_{R}\right)}{x_{R}^{2}}.
\end{split}
\label{DuflofunctLR}
\end{equation}

We can now rewrite the Duflo factor in the canonical basis generated
by $e_{i}\equiv\left(J_{i},N_{i}\right)$ (for which we have canonical
coordinates $k^{i}\equiv\left(\rho^{i},\beta^{i}\right)$). Considering
that an element of the Lie algebra is $x=x_{L}^{i}L_{i}+x_{R}^{i}R_{i}=x_{J}^{i}J_{i}+x_{N}^{i}N_{i}$,
we get 
\begin{equation}
x_{L}^{i}=x_{J}^{i}+ix_{N}^{i},\qquad x_{R}^{i}=x_{J}^{i}-ix_{N}^{i},
\end{equation}
and 
\begin{equation}
j^{\frac{1}{2}}\left(x\right)=4\frac{\left|\sinh\left(\tfrac{1}{2}x_{\zeta}\right)\right|^{2}}{\left|x_{\zeta}^{2}\right|},
\end{equation}
where $x_{\zeta}=\sqrt{\vec{x}_{\zeta}\cdot\vec{x}_{\zeta}}$ and
$x_{\zeta}^{i}=x_{J}^{i}+ix_{N}^{i}$.

Notice that if we rewrite the exponential function on $\g^{*}$ as
\begin{equation}
\exp\left(ik_{I}X^{I}\right)=\exp\left(i\vec{\zeta}\cdot\vec{X}_{L}\right)\exp\left(i\vec{\zeta}^{*}\cdot\vec{X}_{R}\right)\qquad X^{I}\in\g^{*},\label{ExpLR}
\end{equation}
where in this basis the coordinates on $\g^{*}$ are\footnote{This can be seen using the duality relations $\left\langle e_{i},\tilde{e}_{j}\right\rangle =\delta_{ij}$
from which$\left\langle x,X\right\rangle =x_{J}^{i}X_{i}^{J}+x_{N}^{i}X_{i}^{N}=x_{L}^{i}X_{i}^{L}+x_{R}^{i}X_{i}^{R}$.} 
\begin{equation}
X_{i}^{L}=\frac{1}{2}\left(X_{i}^{J}-iX_{i}^{N}\right),\qquad X_{i}^{R}=\frac{1}{2}\left(X_{i}^{J}+iX_{i}^{N}\right),\label{XLXJ-1-1}
\end{equation}
we can also apply the Duflo function directly on the form (\ref{DuflofunctLR}),
i.e., apply the function $j^{\frac{1}{2}}\left(\partial\right)$ to
the exponential (\ref{ExpLR}) to get 
\begin{equation}
\left(j^{\frac{1}{2}}\left(\partial\right)\exp\right)\left(ik^{I}X_{I}\right)=4\frac{\left|\sin\left(\tfrac{1}{2}\zeta\right)\right|^{2}}{\left|\zeta^{2}\right|}\exp\left(ik^{I}X_{I}\right),\label{DufloFunctionExpLR}
\end{equation}
which coincides with (\ref{DufloFunctionExp}).

\subsection{Explicit calculation of the $\star$-product on monomials\label{sec:starProd}}

Considering expressions (\ref{starWaveExpansion}) and (\ref{eq:productE}),
the $\star$-product between $n$ coordinates of $\g^{*}$ can be
evaluated through the formula 
\begin{equation}
X_{i_{1}}\star X_{i_{2}}\star\cdots X_{i_{n}}=\left(-i\right)^{n}\frac{\partial^{n}}{\partial k_{1}^{i_{1}}\partial k_{2}^{j_{2}}\cdots\partial k_{3}^{i_{n}}}\Big|_{k_{1}=k_{2}=\cdots=k_{n}=0}{\cal D}^{-1}\left(e^{i\vec{{\cal B}}\left(k_{1},k_{2},\dots,k_{n}\right)\cdot\hat{X}}\right).\label{StarProdCalc2}
\end{equation}
The term to derivate in the last expression can be rewritten as 
\begin{equation}
{\cal A}\left({\cal B}\left(k_{1},k_{2},\dots,k_{n}\right)\right){\cal S}^{-1}\left(e^{i\vec{k}_{1}\cdot\hat{X}}e^{i\vec{k}_{2}\cdot\hat{X}}\cdots e^{i\vec{k}_{1}\cdot\hat{X}}\right).\label{DufloBCH}
\end{equation}
For the lowest order powers for the$\star$-product we get 
\begin{equation}
X_{i}\star X_{j}=\frac{\left(-i\right)^{2}\partial^{2}}{\partial k_{1}^{i}\partial k_{2}^{j}}\Big|_{k=0}\left({\cal A}_{k(k_{1},k_{2})}^{(2)}\left(k_{1},k_{2}\right)+i^{2}\vec{k}_{1}^{i}\vec{k}_{2}^{j}{\cal S}^{-1}\left(\hat{X}_{i}\hat{X}_{j}\right)\right),\label{starProd2ndO}
\end{equation}
\begin{equation}
\begin{split}X_{i}\star X_{j}\star X_{k}= & \frac{\left(-i\right)^{3}\partial^{3}}{\partial k_{1}^{i}\partial k_{2}^{j}\partial k_{3}^{k}}\Big|_{k=0}\Bigg({\cal A}_{k(k_{1},k_{2},k_{3})}^{(3)}\left(k_{1},k_{2},k_{3}\right)+i^{3}k_{1}^{m}k_{2}^{n}k_{3}^{l}{\cal S}^{-1}\left(\hat{X}_{m}\hat{X}_{n}\hat{X}_{l}\right)\\
 & +i\left({\cal A}_{k(k_{1},k_{2},k_{3})}^{(2)}\left(k_{1},k_{2}\right)k_{3}^{l}+{\cal A}_{k(k_{1},k_{2},k_{3})}^{(2)}\left(k_{1},k_{3}\right)k_{2}^{l}+{\cal A}_{k(k_{1},k_{2},k_{3})}^{(2)}\left(k_{2},k_{3}\right)k_{1}^{l}\right)X_{l}\Bigg),
\end{split}
\label{starProd3rdO}
\end{equation}
where we indicated with ${\cal A}_{k(k_{1},k_{2},\cdots,,k_{n})}^{(i)}\left(k_{a},k_{b},\dots,k_{c}\right)$
the $(i)$-order mixed term in $k_{a},k_{b},\dots,k_{c}$ of the $n$-ple
Duflo factor.

In terms of the complex vector $\vec{\zeta}=\vec{\rho}+i\vec{\beta}$,
defining $\vec{{\cal B}}\left(k_{1},k_{2},\cdots,k_{n}\right)=\vec{\zeta}\left(\zeta_{1},\zeta_{2},\cdots,\zeta_{n}\right)$,
we can calculate the BCH formula from the SL(2,C) representation (\ref{SL(2,C)complexVector})
\begin{equation}
e^{\frac{i}{2}\vec{\zeta}\cdot\sigma}=\cos\left(\tfrac{1}{2}\vec{\zeta}\cdot\sigma\right)\id+i\sin\left(\tfrac{1}{2}\vec{\zeta}\cdot\sigma\right)\hat{\zeta}\cdot\sigma,
\end{equation}
from which follows for the double BCH (the BCH coming from the product
of two exponentials) 
\begin{equation}
\tan\left(\tfrac{1}{2}\zeta\left(\zeta_{1},\zeta_{2}\right)\right)\hat{\zeta}\left(\zeta_{1},\zeta_{2}\right)=\frac{\sin\left(\tfrac{1}{2}\zeta_{1}\right)\cos\left(\tfrac{1}{2}\zeta_{2}\right)\hat{\zeta}_{1}+\cos\left(\tfrac{1}{2}\zeta_{1}\right)\sin\left(\tfrac{1}{2}\zeta_{2}\right)\hat{\zeta}_{2}-\sin\left(\tfrac{1}{2}\zeta_{1}\right)\sin\left(\tfrac{1}{2}\zeta_{2}\right)\hat{\zeta}_{1}\wedge\hat{\zeta}_{2}}{\cos\left(\tfrac{1}{2}\zeta_{1}\right)\cos\left(\tfrac{1}{2}\zeta_{2}\right)-\sin\left(\tfrac{1}{2}\zeta_{1}\right)\sin\left(\tfrac{1}{2}\zeta_{2}\right)\hat{\zeta}_{1}\cdot\hat{\zeta}_{2}}.\label{BCHSL2C}
\end{equation}
The Duflo factor (\ref{EstarE}) (or inverse (\ref{DufloFunctionExp}))
can be expanded as ($\left\{ k\right\} =\left\{ \zeta\right\} $ or
$\left\{ \rho,\beta\right\} $) 
\begin{equation}
{\cal A}=1+\frac{1}{24}\left(\zeta^{2}+\zeta^{*}{}^{2}\right)+O\left(k^{4}\right)=1+\frac{1}{12}\left(\rho^{2}-\beta^{2}\right)+O\left(k^{4}\right).\label{DufloExpansion}
\end{equation}
Since the BCH (\ref{BCHSL2C}) for $\vec{\zeta}$ is at least linear
in $\zeta_{1}$ and $\zeta_{2}$, it is enough to consider the expansion
of the BCH up to 2nd order in the coordinates $\left\{ k\right\} $
\begin{equation}
\vec{\zeta}\left(\zeta_{1},\zeta_{2}\right)\simeq\vec{\zeta}_{1}+\vec{\zeta}_{2}-\frac{1}{2}\vec{\zeta}_{1}\wedge\vec{\zeta}_{2},
\end{equation}
or 
\begin{equation}
\begin{gathered}\vec{\rho}\left(\rho_{1},\rho_{2}\right)\simeq\vec{\rho}_{1}+\vec{\rho}_{2}-\frac{1}{2}\left(\vec{\rho}_{1}\wedge\vec{\rho}_{2}-\vec{\beta}_{1}\wedge\vec{\beta}_{2}\right),\\
\vec{\beta}\left(\beta_{1},\beta_{2}\right)\simeq\vec{\beta}_{1}+\vec{\beta}_{2}-\frac{1}{2}\left(\vec{\rho}_{1}\wedge\vec{\beta}_{2}+\vec{\beta}_{1}\wedge\vec{\rho}_{2}\right).
\end{gathered}
\label{BCH2ndO}
\end{equation}
For the triple BCH we can use the associativity\footnote{The associativity of the BCH comes from the associativity of the group
product.} property ${\cal B}\left(\zeta_{1},\zeta_{2},\zeta_{3}\right)={\cal B}\left(\zeta_{1},{\cal B}\left(\zeta_{2},\zeta_{3}\right)\right)$,
from which we get, up to 2nd order 
\begin{equation}
\begin{gathered}\vec{\rho}\left(\rho_{1},\rho_{2},\rho_{2}\right)\simeq\vec{\rho}_{1}+\vec{\rho}_{2}+\vec{\rho}_{3}-\frac{1}{2}\left(\vec{\rho}_{1}\wedge\vec{\rho}_{2}+\vec{\rho}_{1}\wedge\vec{\rho}_{3}+\vec{\rho}_{2}\wedge\vec{\rho}_{3}-\vec{\beta}_{1}\wedge\vec{\beta}_{2}-\vec{\beta}_{1}\wedge\vec{\beta}_{3}-\vec{\beta}_{2}\wedge\vec{\beta}_{3}\right),\\
\vec{\beta}\left(\beta_{1},\beta_{2},\beta_{2}\right)\simeq\vec{\beta}_{1}+\vec{\beta}_{2}+\vec{\beta}_{3}-\frac{1}{2}\left(\vec{\rho}_{1}\wedge\vec{\beta}_{2}+\vec{\rho}_{1}\wedge\vec{\beta}_{3}+\vec{\rho}_{2}\wedge\vec{\beta}_{3}+\vec{\beta}_{1}\wedge\vec{\rho}_{2}+\vec{\beta}_{1}\wedge\vec{\rho}_{3}+\vec{\beta}_{2}\wedge\vec{\rho}_{3}\right).
\end{gathered}
\label{3pleBCH2ndO}
\end{equation}
The symmetrization map (\ref{symmMap}) on 2nd and 3rd order monomials
is such that 
\begin{equation}
{\cal S}^{-1}\left(\hat{X}_{i}\hat{X}_{j}\right)=X_{i}X_{j}+\frac{1}{2}{\cal S}^{-1}\left(\left[\hat{X}_{i},\hat{X}_{j}\right]\right),\label{sym2ndO}
\end{equation}
\begin{equation}
\begin{split}S^{-1}\left(\hat{X}_{i}\hat{X}_{j}\hat{X}_{k}\right)= & X_{i}X_{j}X_{k}+\frac{1}{2}\left(X_{i}{\cal S}^{-1}\left(\left[\hat{X}_{j},\hat{X}_{k}\right]\right)+X_{j}S^{-1}\left(\left[\hat{X}_{i},\hat{X}_{k}\right]\right)+X_{k}{\cal S}^{-1}\left(\left[\hat{X}_{i},\hat{X}_{j}\right]\right)\right)\\
 & +\frac{1}{12}{\cal S}^{-1}\left(3\left[\hat{X}_{i},\left[\hat{X}_{j},\hat{X}_{k}\right]\right]-\left[\hat{X}_{j},\left[\hat{X}_{i},\hat{X}_{k}\right]\right]-\left[\hat{X}_{k},\left[\hat{X}_{i},\hat{X}_{j}\right]\right]\right).
\end{split}
\label{sym3rdO}
\end{equation}
Substituting (\ref{BCH2ndO}), (\ref{3pleBCH2ndO}) in (\ref{DufloExpansion}),
and using the commutation relations (\ref{QalgebraX}) we can finally
calculate from (\ref{starProd2ndO}) and (\ref{starProd3rdO}). We
here report the explicit expression of third order monomials for the
Duflo-map $\star$-product (repeated indexes are summed): 
\begin{equation}
\begin{gathered}X_{i}^{J}\star X_{j}^{J}\star X_{k}^{J}=X_{i}^{J}X_{j}^{J}X_{k}^{J}+\frac{i}{2}\left(\epsilon_{ij}^{\ l}X_{k}^{J}X_{l}^{J}+\epsilon_{ik}^{\ l}X_{j}^{J}X_{l}^{J}+\epsilon_{jk}^{\ l}X_{i}^{J}X_{l}^{J}\right)-\frac{1}{2}\delta_{ik}X_{j}^{J}-\frac{i}{12}\epsilon_{ijk},\\
X_{i}^{J}\star X_{j}^{J}\star X_{k}^{N}=X_{i}^{J}X_{j}^{J}X_{k}^{N}+\frac{i}{2}\left(\epsilon_{ij}^{\ l}X_{k}^{N}X_{l}^{J}+\epsilon_{ik}^{\ l}X_{j}^{J}X_{l}^{N}+\epsilon_{jk}^{\ l}X_{i}^{J}X_{l}^{N}\right)+\frac{1}{6}\left(\delta_{jk}X_{i}^{N}-2\delta_{ik}X_{j}^{N}\right),\\
X_{i}^{N}\star X_{j}^{N}\star X_{k}^{N}=X_{i}^{N}X_{j}^{N}X_{k}^{N}-\frac{i}{2}\left(\epsilon_{ij}^{\ l}X_{k}^{N}X_{l}^{J}+\epsilon_{ik}^{\ l}X_{j}^{N}X_{l}^{J}+\epsilon_{jk}^{\ l}X_{i}^{N}X_{l}^{J}\right)+\frac{1}{2}\left(\delta_{ik}X_{j}^{N}\right),\\
X_{i}^{N}\star X_{j}^{N}\star X_{k}^{J}=X_{i}^{N}X_{j}^{N}X_{k}^{J}-\frac{i}{2}\left(\epsilon_{ij}^{\ l}X_{k}^{J}X_{l}^{J}-\epsilon_{ik}^{\ l}X_{j}^{N}X_{l}^{N}-\epsilon_{jk}^{\ l}X_{i}^{N}X_{l}^{N}\right)+\frac{1}{6}\left(2\delta_{ik}X_{j}^{J}-\delta_{jk}X_{i}^{J}\right)+\frac{i}{12}\epsilon_{ijk}.
\end{gathered}
\end{equation}

From the properties of the Duflo map it is also easy to prove that
the Duflo star product preserves the algebra of $\text{Sym}\left(\g\right)^{\g}$.
One can check explicitly for instance that give the quadratic Casimir
$\hat{{\cal C}}=\hat{X}_{J}^{2}-\hat{X}_{N}^{2}$ of the quantum algebra,
such that 
\begin{equation}
{\cal C}_{\star}={\cal D}^{-1}\left(\hat{{\cal C}}\right)=X_{i}^{J}\star X_{i}^{J}-X_{i}^{N}\star X_{i}^{N}=\vec{X}_{J}^{2}-\vec{X}_{N}^{2}-1,\label{CasimirStar}
\end{equation}
\begin{equation}
\hat{{\cal C}}\cdot\hat{{\cal C}}={\cal D}\left({\cal C}_{\star}\right)\cdot{\cal D}\left({\cal C}_{\star}\right)={\cal D}\left({\cal C}_{\star}^{2}\right),
\end{equation}
i.e. 
\begin{equation}
{\cal C}_{\star}\star{\cal C}_{\star}={\cal C}_{\star}^{2}\label{DufloCasimir}
\end{equation}

\section{Some properties of the homogeneous space ${\cal H}_{3}\approx\text{SL}\left(2,\C\right)/\text{SU}(2)$}

\subsection{Hyperbolic rotation angle \label{sec:HyperAngle}}

We show in this subsection the relation between the Cartan parametrization
of Sec.~\ref{sec:SL2C/SU2} and the parametrization in terms of ``hyperbolic
rotations'' used in~\cite{JunkerBohm,JunkerInomata}. These can
be considered as the generalization to hyperbolic space of Euler angles
rotations. The only change respect to the Cartan splitting of Sec.~\ref{sec:SL2C/SU2}
is in the ``boost'' sector $\text{SL}(2,\C)/\text{SU}(2)$ (or SO(3,1)/SO(3)).
In this case an element of $\text{SL}(2,\C)$ (or SO(3,1)) is split
as 
\begin{equation}
g=\kk h,
\end{equation}
where $h\in\text{SU}(2)$ (or SO(3)), and $\kk\in\text{SL}(2,\C)/\text{SU}(2)$
(or SO(3,1)/SO(3)) such that 
\begin{equation}
\kk=h_{1}\left(\alpha_{1}\right)h_{3}\left(\alpha_{3}\right)\kk_{3}\left(\Theta\right),
\end{equation}
where $h_{1}\left(\alpha_{1}\right)$ and $h_{3}\left(\alpha_{3}\right)$
are Euclidean rotations around respectively the 1 and 3 axes, while
$\kk_{3}\left(\Theta\right)$ is an hyperbolic rotation (a boost)
along the 3 axis: 
\begin{equation}
\kk_{3}\left(\Theta\right)=\left\{ \begin{gathered}\cosh\left(\tfrac{1}{2}\Theta\right)\id_{2}-\sinh\left(\tfrac{1}{2}\Theta\right)\sigma_{3}\qquad\text{in}\ \text{SL}(2,\C)/\text{SU}(2),\\
\\
\left(\begin{array}{cccc}
\cosh\left(\Theta\right) & 0 & 0 & \sinh\left(\Theta\right)\\
0 & 1 & 0 & 0\\
0 & 0 & 1 & 0\\
\sinh\left(\Theta\right) & 0 & 0 & \cosh\left(\Theta\right)
\end{array}\right)\qquad\text{in}\ \text{SO}(3,1)/\text{SO}(3).
\end{gathered}
\right.
\end{equation}
The action of $\kk$ on the origin $q_{a}=\left(1,\vec{0}\right)$
of ${\cal H}_{3}$ gives a generic point of $q=\kk q_{a}\kk^{\dagger}$
($q=\kk q_{a}$ for $\text{SO}(3,1)/\text{SO}(3)$) of ${\cal H}_{3}$
as 
\begin{equation}
\begin{gathered}q_{0}=\cosh\left(\Theta\right),\\
q_{1}=\sinh\left(\Theta\right)\sin\left(\alpha_{3}\right)\sin\left(\alpha_{1}\right),\\
q_{2}=\sinh\left(\Theta\right)\sin\left(\alpha_{3}\right)\cos\left(\alpha_{1}\right),\\
q_{3}=\sinh\left(\Theta\right)\cos\left(\alpha_{3}\right).
\end{gathered}
\end{equation}
By comparison with Eq.~(\ref{HyperPoint}) we find that 
\begin{equation}
\begin{gathered}\Theta=\mathfrak{b},\\
\hat{\mathfrak{b}}=\left(\sin\left(\alpha_{3}\right)\sin\left(\alpha_{1}\right),\sin\left(\alpha_{3}\right)\cos\left(\alpha_{1}\right),\cos\left(\alpha_{3}\right)\right).
\end{gathered}
\end{equation}
Moreover, considering the Jacobian 
\begin{equation}
d^{3}\vec{\mathfrak{b}}=\Theta^{2}\sin\left(\alpha_{1}\right)d\Theta d\alpha_{1}d\alpha_{3},
\end{equation}
the measure~\ref{HaarCartan} becomes 
\begin{equation}
dg=\sinh^{2}\left(\Theta\right)\sin\left(\alpha_{1}\right)d\Theta d\alpha_{1}d\alpha_{3}dh.
\end{equation}

\subsection{Projection to ${\cal H}_{3}$ of the propagator\label{sec:ProjectionH3}}

The propagator in ${\cal H}_{3}$ can be rewritten as 
\begin{equation}
\begin{split}K\left(q'',q';t\right)= & \braket{q''|U(t)|q'}\\
= & \int_{\text{SL}(2,\C)}dg'\int_{\text{SL}(2,\C)}dg''\ \braket{q''|g''}\braket{g''|U(t)|g'}\braket{g'|q'}
\end{split}
\end{equation}
Defining 
\begin{equation}
\braket{g'|q'}\braket{q''|g''}=\delta\left(q'{k'}^{-1}\right)\delta\left(q''{k''}^{-1}\right)\delta\left(h'{h''}^{-1}\right)
\end{equation}
for the Cartan splitting (\ref{cartan}) $g'=k'h'$, $g''=k''h''$,
we get 
\begin{equation}
\begin{split}K\left(q'',q';t\right)= & \int_{{\cal H}_{3}}dk'\int_{\text{SU}(2)}dh'\int_{{\cal H}_{3}}dk''\int_{\text{SU}(2)}dh''\ \delta\left(q'{k'}^{-1}\right)\delta\left(q''{k''}^{-1}\right)\delta\left(h'{h''}^{-1}\right)\braket{g''|U(t)|g'}\\
= & \int_{\text{SU}(2)}dh\ \braket{q''h|U(t)|q'h}
\end{split}
\end{equation}
i.e. 
\begin{equation}
K\left(q'',q';t\right)=\int_{\text{SU}(2)}dh\ K_{G}\left(q''h,q'h;t\right)=\int_{\text{SU}(2)}dh\ K_{G}\left(g=g''{g'}^{-1};t\right)
\end{equation}
where we also used the invariance of the SU(2) Haar measure $d\left(hh'\right)=dh$.

\end{document}